\documentclass[prd,showpacs,amsmath,amssymb,nofootinbib,superscriptaddress]{revtex4}
\usepackage{array,mathtools,amssymb,booktabs,multirow,diagbox,hhline}
\pdfoutput=1

\usepackage{graphicx}
\usepackage{dcolumn}
\usepackage{bm}
\usepackage{bbold}
\usepackage{subfig}
\usepackage{graphicx}

\newcommand{\Od}{{\cal O}}
\newcommand{\pint}{\int \frac{d^3 \vec{p}}{(2\pi)^3}}
\newcommand{\im}{\mbox{Im}\,}
\newcommand{\re}{\mbox{Re}\,}
\newcommand{\intT}{\int_0^\beta d\tau \int d^3 \vec{x}}

\newcommand{\mean}[1]{\left\langle{#1}\right\rangle}
\newcommand{\quarkcorl}{\langle {\cal T} (\bar \psi_l \psi_l (x)  \bar \psi_l \psi_l (0)\rangle}

\newcommand{\conds}{\langle \bar s s \rangle}
\newcommand{\condl}{\mean{\bar q q}_l}

\begin{document}

\title{The role of the thermal $f_0(500)$ in chiral symmetry restoration}
\author{S. Ferreres-Sol\'e}
\email{ferreres.sole@gmail.com}
\affiliation{NIKHEF,
Science Park 105,
NL-1098 XG,  Amsterdam
Netherlands}
\author{A. G\'omez Nicola}
\email{gomez@ucm.es}
\affiliation{Departamento de F\'{\i}sica
Te\'orica and IPARCOS. Univ. Complutense. 28040 Madrid. Spain}
\author{A. Vioque-Rodr\'iguez}
\email{avioque@ucm.es}
\affiliation{Departamento de F\'{\i}sica
Te\'orica and IPARCOS. Univ. Complutense. 28040 Madrid. Spain}

\begin{abstract}
We show that the $\sigma/f_0(500)$ state with finite-temperature $T$ corrections to its spectral properties included, plays an essential role for the description of the scalar susceptibility $\chi_S$, signaling chiral symmetry restoration. First, we use the $O(4)$ Linear Sigma Model as a testbed to derive the connection between $\chi_S$ and the $\sigma$ propagator and to check the validity and reliability of the  approach where $\chi_S$ is saturated by the $\sigma/f_0(500)$ inverse self-energy, which we calculate at finite $T$ to one loop.  A more accurate phenomenological description is achieved by considering  the saturation approach as given by the thermal $f_0(500)$ state generated in Unitarized Chiral Perturbation Theory. Such approach   allows to describe fairly well recent lattice data within the uncertainty range given by the UChPT parameters. Finally, we compare the UChPT saturated description  with one based on the  Hadron Resonance Gas, for which the hadron mass dependences are extracted from recent theoretical analysis. Several fits to lattice data are performed, which confirm the validity of the thermal $f_0(500)$ saturated approach and hence  the importance of that thermal state for chiral symmetry restoration. 
\end{abstract}

\maketitle

\section{Introduction}
\label{sec:intro}

Chiral symmetry restoration and its nature are surely among  the key problems for our present understanding of the QCD phase diagram. It is well established that in the physical case of $N_f=2+1$ flavours with $m_l\ll m_s$  quark masses, the chiral transition is a  crossover at a transition temperature of about $T_c\sim 155-160$ MeV for vanishing baryon density~\cite{Aoki:2009sc,Borsanyi:2010bp,Bazavov:2011nk,Buchoff:2013nra}. The ideal chiral restoration phase transition is reached only for $N_f=2$ and $m_l =0$, while in the physical case it is approached in the light chiral limit $m_l\rightarrow 0^+$ \cite{Pisarski:1983ms}.

The main signals of a chiral restoration crossover are, on the one hand, the decreasing behaviour of the quark condensate $\condl=\mean{\bar\psi_l\psi_l}$, where $\psi_l=\left(\begin{array}{c}u\\d\end{array}\right)$ is the light quark doublet, and, on the other hand, a peak in the scalar susceptibility $\chi_S(T)$, where

\begin{eqnarray}
\condl(T)&=&\partial z(T)/\partial m_l, 
\label{condef}\\
\chi_S (T)&=&-\frac{\partial}{\partial m_l} \condl(T)=\int_T {d^4x \left[\quarkcorl-\condl^2(T)\right]},
\label{susdef}
\end{eqnarray}
 $\displaystyle \int_T dx\equiv \intT$ at finite temperature $T=1/\beta$, we are considering the isospin limit $m_u=m_d=m_l $ and $\mean{\cdot}$ denotes Euclidean finite-$T$ correlators. The  free energy density in the above equations is  $z(T)=-\lim_{V\rightarrow\infty}(\beta V)^{-1}\log Z$ at finite temperature $T$ with vanishing chemical potentials, with $Z$ the QCD partition function or its hadronic realization through an effective theory.  Thus, the scalar susceptibility $\chi_S$ in  \eqref{susdef} should peak at the chiral transition, or diverge in the light chiral limit for $N_f=2$ \cite{Smilga:1995qf,Ejiri:2009ac} and this is indeed reflected in lattice data \cite{Aoki:2009sc} where the peak of $\chi_S$ confirms the crossover nature of the transition in the physical limit. 

From the theoretical side, it is important  to provide reliable approximations which could describe the expected behaviour for the quark condensate and the scalar susceptibility and eventually be used to fit lattice data. The most widely used approach in this context has been the Hadron Resonance Gas (HRG) approximation 
\cite{Hagedorn:1968zz,Karsch:2003zq,Karsch:2003vd,Tawfik:2005qh,Leupold:2006ih,Huovinen:2009yb,Megias:2012kb,Jankowski:2012ms}. Within the HRG, the pressure of the system is described as a collection of free resonances, including in principle all hadron states quoted by the Particle Data Group (PDG) \cite{Tanabashi:2018oca} up to a given energy above which Boltzmann suppression is effective. Thus, hadron interactions are meant to be encoded  through their corresponding resonant channels, and the width of the resonant states is usually neglected.   This approximation works quite  well below $T_c$, where it is meant to be valid, although qualitatively it does not reproduce the inflection point expected for the quark condensate, nor, as we will see in detail here, the peak of the scalar susceptibility in the crossover regime. 

Calculations of the thermodynamics including interactions among hadrons encompass the Chiral Perturbation Theory (ChPT) description of the light meson gas \cite{Gerber:1988tt}, which provides a model-independent and consistent treatment. Although  restricted to low temperatures, it captures well the contribution of the lightest states.  An alternative  is the virial approach, where  interactions are incorporated within a small fugacity expansion. Both ChPT and the virial approach predict similar results  consistent with chiral symmetry restoration \cite{GomezNicola:2012uc}, although higher order states are needed, as provided by the HRG, to obtain results compatible with the lattice.

A relevant issue, not often considered, is the importance of the  interactions among  the thermal bath components, which ultimately  give rise to a  temperature dependence of the hadron spectral parameters (mass and width, generally speaking). As a particular example of relevance for the present work,  the ChPT analysis of pion scattering at finite temperature combined with unitarity arguments  \cite{GomezNicola:2002tn,Dobado:2002xf} allows to obtain the temperature dependence of the $\rho(770)$ and $f_0(500)$ poles in the second Riemann sheet (2RS). In the $\rho$ case, the results are compatible with the observed widening in dilepton spectrum, parametrized through the pion form factor \cite{GomezNicola:2004gg}, but the more important consequence for our present work concerns the $f_0(500)$ channel, since the analysis in \cite{Nicola:2013vma} shows that the scalar susceptibility saturated by such thermal $f_0(500)$ state has a maximum very close to the expected transition point, which is quite remarkable given the approximations used.

In this work we will explore in detail some phenomenological and theoretical aspects related to  the scalar susceptibility description through the $f_0(500)$ thermal state. Our aim and motivation are to investigate to what extent the thermal $f_0(500)$ must be taken into account when describing observables regarding the chiral transition.  The scalar susceptibility is the candidate for which the influence of such state is meant to be dominant, since, as we will explain below in detail, it should scale roughly as the inverse thermal mass squared of the scalar propagator, for which  the $f_0(500)$ gives the lightest contribution.   

Our analysis will proceed along the following lines: the formal connection of the $f_0(500)$ with the scalar susceptibility will be discussed in section \ref{sec:unitf0}, where we will focus the discussion on the $O(4)$ Linear Sigma Model (LSM) as an example of a theory including explicitly the $\sigma$ degree of freedom, where the connection between the scalar susceptibility and the $\sigma$ self-energy will be analyzed, and on Unitarized ChPT (UChPT), which  provides a more accurate description of the $f_0(500)$ state at $T=0$ without the need of such state in the lagrangian. In both cases we will see that the saturated approach provides a description closer to lattice data, the UChPT approach reproducing the expected crossover peak unlike the LSM one.  Our next step (section \ref{sec:uchpt}) will be to check the robustness of the unitarized saturation approach and its capability to describe lattice data without further approximations. For that purpose, we will study the sensibility of the model to the uncertainties in the Low-Energy Constants (LEC) of ChPT and to the requirements imposed by the unitarization method, such as unitarity, analiticity and a good determination of the $T=0$ pole. In section \ref{sec:hrg}, we will present a HRG analysis of the scalar susceptibility, which as far as we know has not been studied so far. Finally, in section \ref{sec:fits} we will perform detailed fits of lattice data to the unitarized model, comparing it with the HRG description. Our main conclusions are summarized in section \ref{sec:conc}.  

The present analysis will confirm the importance of considering thermal (or generally in-medium) interactions to describe certain hadron gas observables and the relevance in particular of the $f_0(500)$ thermal sate,  opening up interesting possibilities for future theoretical and lattice studies.

\section{The  thermal $\sigma/f_0(500)$ and the scalar susceptibility}
\label{sec:unitf0}

In this section we will discuss the connection of the scalar susceptibility, defined in \eqref{susdef}, with the lightest scalar meson state, the $\sigma$ or $f_0(500)$, which is indexed in the Particle Data Book as a broad resonance arising dominantly in $\pi\pi$ scattering \cite{Tanabashi:2018oca}. The  $f_0(500)$  and its main properties have been recently reviewed in \cite{Pelaez:2015qba}. This connection will be studied first within the framework of the LSM as a testbed which will allow to check the different approximations used. Secondly, within   UChPT we will consider a saturated approach in terms of the thermal $f_0(500)$ generated as a pole in $\pi\pi$ scattering.

\subsection{Linear Sigma Model description}
\label{sec:lsm}
First, for clarifying purposes, let us consider a meson field theory realization of low-energy QCD where there is an explicit realization of the scalar $\sigma$ as a fundamental field in the lagrangian. A particular example  is the LSM or $O(4)$ vector model \cite{GellMann:1960np}, which exhibits chiral symmetry restoration properties  \cite{Bochkarev:1995gi,Ayala:2000px}. For our present discussion it will be enough to consider the light meson sector lagrangian of this model in terms of sigma $\sigma$ and pion $\pi^a$ fields:

\begin{equation}
{\cal L}_{LSM}= \frac{1}{2}\partial_\mu\Phi^T\partial^\mu \Phi-\frac{\lambda}{4}\left[\Phi^T\Phi-v_0^2\right]^2+h\sigma,
\label{lsm1}
\end{equation}
with $\Phi^T=(\sigma,\vec{\pi})$ and we have chosen, as usual, the $\sigma$ direction to break the symmetry $O(4)\rightarrow O(3)$. The $h$ term breaks explicitly the chiral symmetry, with $h$  proportional to the pion mass squared, whereas the potential minima at $\Phi^2=v^2\neq 0$  implement spontaneous chiral symmetry breaking. For $h=0$, $v=v_0$ (chiral limit) whereas for $h\neq 0$, the value of $v$  is determined by the minimum of the potential $V(\sigma)$ as

\begin{equation}
h=\lambda v (v^2-v_0^2).
\label{lsmv}
\end{equation}

The $T=0$ standard procedure is  to shift the field  as $\tilde\sigma=\sigma-v$, so that  $\mean{\tilde\sigma}=0$ to leading order. At $T\neq 0$ however, $\mean{\sigma} (T)\equiv v(T) \neq v$ so if one decides to use the same shifted $\tilde \sigma$ field, as done for instance in \cite{Ayala:2000px}, it should be taken into account that $\mean{\tilde\sigma}(T)=v(T)-v\neq 0$, which in particular implies that one-particle reducible (1PR) diagrams enter in the calculation of correlators, as in the case of the $\tilde\sigma$ propagator. An alternative, followed  in \cite{Bochkarev:1995gi}, is to use instead the shifted field $\hat \sigma=\sigma-v(T)$, so that $\mean{\hat\sigma}=0$. The temperature dependence of $v(T)$ can be determined within the LSM, for instance from the mean field approach to leading order in $\lambda$ \cite{Bochkarev:1995gi}. 

With the first prescription,  the lagrangian \eqref{lsm1} becomes, in terms of the shifted $\tilde\sigma$ field,

\begin{equation}
{\cal L}_{LSM}= \frac{1}{2}\left(\partial_\mu\tilde\sigma\partial^\mu\tilde\sigma+\partial_\mu\pi^a\partial^\mu\pi^a-M_{0\sigma}^2\tilde\sigma^2-M_{0\pi}^2\pi_a\pi^a\right)-\frac{\lambda}{4}\left(\tilde\sigma^2+\pi_a\pi^a\right)^2-\lambda v \tilde\sigma\left(\tilde\sigma^2+\pi_a\pi^a\right)-\frac{1}{4\lambda}M_{0\pi}^4+v^2M_{0\pi}^2,
\label{lsm2}
\end{equation}
where $M_{0\pi}$ and $M_{0\sigma}$   are the tree-level pion and sigma masses 

\begin{equation}
M_{0\pi}^2=\frac{h}{v}=\lambda(v^2-v_0^2) \quad , \quad M_{0\sigma}^2=M_{0\pi}^2+2\lambda v^2,
\label{massesLSM}
\end{equation}
and where, in order to comply with low-energy theorems, or ChPT to leading order \cite{Gasser:1983yg} ($\lambda\rightarrow\infty$,  $\Phi^T \Phi=v_0^2$ in \eqref{lsm1}) we have $v_0=F$,  the pion decay constant in the chiral limit, so that $v=v_0(1+\Od(M_\pi^2/M_\sigma^2))=F(1+\Od(M_\pi^2/M_\sigma^2))=F_\pi(1+\Od(M_\pi^2/M_\sigma^2))$ with $F_\pi\simeq$ 92.3 MeV. In addition, we write as is customary $M_{0\pi}^2=2B_0 m_l$ so that to leading order in the chiral low-energy expansion, the Gell-Mann-Oakes-Renner relation $\condl (T=0)=-2B_0 F^2(1+\Od(M_\pi^2/M_\sigma^2))$ holds.

Thus, the quark condensate  \eqref{condef} and  the scalar susceptibility \eqref{susdef} of the LSM can be written as follows: 

\begin{eqnarray}
\condl (T)&=&-\frac{dh}{dm_l}v(T), \label{condlsm}\\
\chi_S (T)&=&\left(\dfrac{d^2 h}{d m_l^2}\right)v(T)+\left(\dfrac{d h}{d m_l}\right)^2 \int_T dx \hspace{0.1cm} 
\left\{\langle \mathcal{T}\tilde\sigma\left(x\right)\tilde\sigma\left(0\right)\rangle-\langle\tilde\sigma\rangle^2 (T)\right\}.
\end{eqnarray}

 The subtraction of $\langle\tilde\sigma\rangle^2$ in $\chi_S$ above ensures that the  self-energy can be written in terms of standard connected Feynman diagrams (including 1PR contributions) and is free of contact divergences proportional to $\delta^{(4)}(k=0)$.  Nevertheless, we will work in the Dimensional Regularization (DR) scheme, so that $\delta^{(D)}(0)$ terms formally vanish \cite{Leibbrandt:1975dj}. Thus, we can write:

\begin{eqnarray}
\chi_S (T)&=&\left(\dfrac{d^2 h}{d m_l^2}\right)v(T)+\left(\dfrac{d h}{d m_l}\right)^2 \Delta_\sigma (k=0;T),
\label{susmod}
\end{eqnarray}

where

\begin{eqnarray}
\Delta_\sigma (k;T)
=\frac{1}{k^2+M_{0\sigma}^2+\Sigma(k_0,\vec{k};T)}
\end{eqnarray}
is the Euclidean propagator of the $\tilde\sigma$ field and  $\Sigma(k_0,\vec{k};T)$ is the self-energy, which in the thermal field theory framework  depends separately on the space and time components of the four-momentum $k$ \cite{galekapustabook}.

The coefficients of $v(T)$ and $\Delta_\sigma (s=0;T)$ in \eqref{susmod}  can be written in terms of $v$, $M_{0\sigma}$ and $M_{0\pi}$, using  \eqref{massesLSM}, as

\begin{eqnarray}
\dfrac{dh}{dm_l}&=&2B_0v\left(\dfrac{M_{0\sigma}^2}{M_{0\sigma}^2-M_{0\pi}^2}\right),\nonumber\\
\dfrac{d^2h}{dm_l^2}&=&4B_0^2 v \frac{2M_{0\sigma}^2-3M_{0\pi}^2}{\left(M_{0\sigma}^2-M_{0\pi}^2\right)^2}.
\end{eqnarray}

The result \eqref{susmod} allows to relate the scalar susceptibility with the propagator of the scalar field. Note that the first term of that equation is meant to be negligible near chiral restoration since it vanishes proportionally to the light quark condensate. Another argument that leads to the same conclusion is the following: around the transition region, $\chi_S$ tends to become degenerate with the pseudoscalar pion susceptibility $\chi_\pi$ ~\cite{Hatsuda:1985eb,Bernard:1987im,Krippa:2000jh}, confirmed in lattice simulations \cite{Buchoff:2013nra}. On the other hand, a Ward Identity allows to write  $\chi_\pi=-\condl/m_l$ at any temperature \cite{Broadhurst:1974ng,Bochicchio:1985xa,Nicola:2013vma,Buchoff:2013nra,Nicola:2016jlj}. Therefore, near the transition, replacing $v(T)=-(dh/dm_l)^{-1} \condl(T)=-m_l\-(dh/dm_l)^{-1}  \chi_\pi (T)\simeq -m_l\-(dh/dm_l)^{-1}   \chi_S (T)$ in the first term in the r.h.s. of \eqref{susmod}, that term is $\Od\left(M_{0\pi}^2/M_{0\sigma}^2\right)$ suppressed with respect to $\chi_S(T)$ in the l.h.s. of \eqref{susmod}.

Therefore, near the transition, the scalar susceptibility is proportional to the $s=0$ euclidean scalar propagator, and its temperature behaviour in that region is  therefore dominated by the inverse self-energy of the lightest state, which in this case is the $\sigma$ of  the LSM:

\begin{equation}
\chi_S (T)\simeq 4B_0^2v^2\left(\dfrac{M_{0\sigma}^2}{M_{0\sigma}^2-M_{0\pi}^2}\right)^2 \Delta_{\sigma}(k=0;T)\Rightarrow \frac{\chi_S (T)}{\chi_S (0)}\simeq \dfrac{M_{0\sigma}^2+\Sigma\left(k=0;T=0\right)}{M_{0\sigma}^2+\Sigma\left(k=0;T\right)}.
\label{susgreen}
\end{equation}

We will refer to the above result as the saturated LSM approach, where the scalar susceptibility is approximately described as the inverse of the self-energy of the lightest scalar state. A similar approach will be carried out and studied extensively in sections \ref{sec:uchpt} and \ref{sec:fits} for the UChPT framework.  In the latter approach, the self-energy contribution is taken as the real part of the pole position in the 2RS of the complex $s=k^2$ plane for the pion scattering amplitude (in the center of momentum frame $\vec{k}=\vec{0}$).  We will then use the present LSM analysis to study the above saturation approach, in particular to compare it with a strictly perturbative result for $\chi_S$ (see details below) so that we can use it as a testbed for the UChPT description, more realistic concerning the $f_0(500)$ pole determination. 

The first step  would be then to provide a well-defined calculation of the $\sigma$ self-energy in the LSM including finite-temperature corrections. For clarity, we will stick to the perturbative approach in $\lambda$. Although, as we are about to check, the real values of $\lambda$ needed to achieve reasonable phenomenological results are quite large \cite{Pelaez:2015qba}, the one-loop corrections to the self-energy lie around a 15\% at $T=0$ \cite{Pelaez:2015qba}. Besides, as commented above, our main goal within the LSM is not to provide a reliable phenomenological description, but to provide a better understanding of the different  approximations for $\chi_S$ performed in this work, at least parametrically in $\lambda$.

The one-loop diagrams contributing to the sigma self-energy are given in Fig. \ref{diagramLSM}. It is important to remark that  a consistent perturbative expansion requires that $M_{0\sigma}^2,M_{0\pi}^2$ remain of $\Od(1)$ in the $\lambda$ expansion so that $k^2+M_\sigma^2$ remains the leading order of the inverse propagator. Hence, using \eqref{massesLSM}, all the one-loop self-energy contributions remain of $\Od(\lambda)$. Namely, the contributions of every diagram in Fig. \ref{diagramLSM} read \cite{Ayala:2000px}:

\begin{figure}
	\centering
	\subfloat[]{
		\label{LSMa}
		\includegraphics[width=0.2\textwidth]{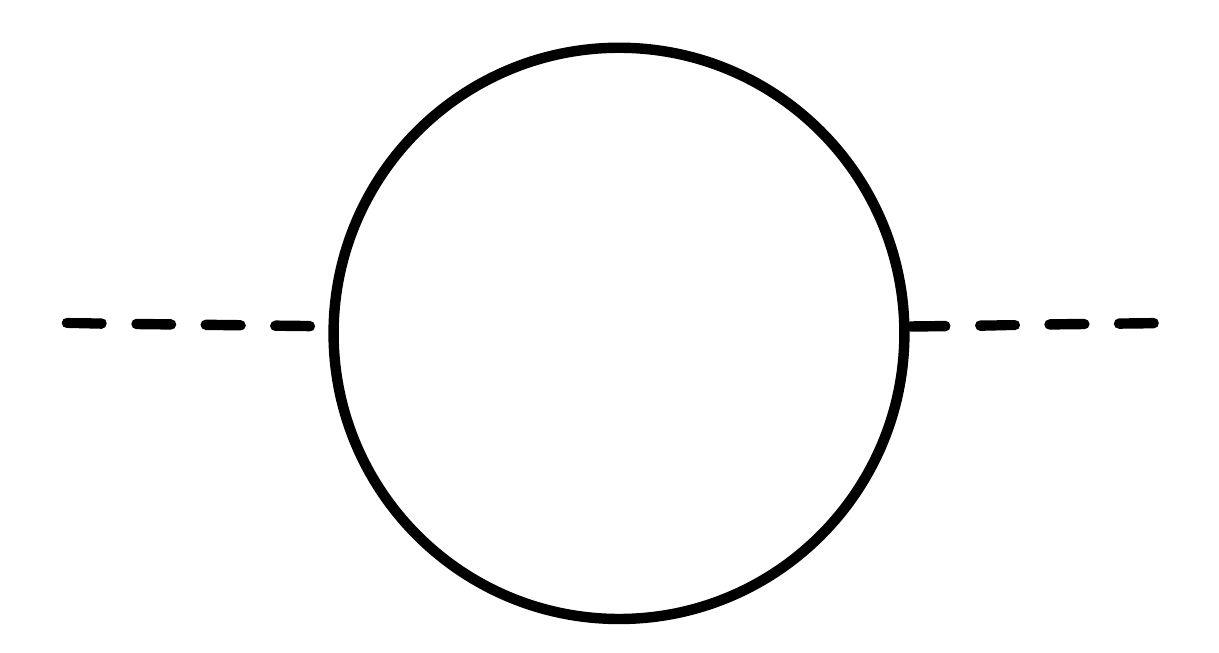}}
	\hspace{1cm}
	\subfloat[]{
		\label{LSMb}
		\includegraphics[width=0.2\textwidth]{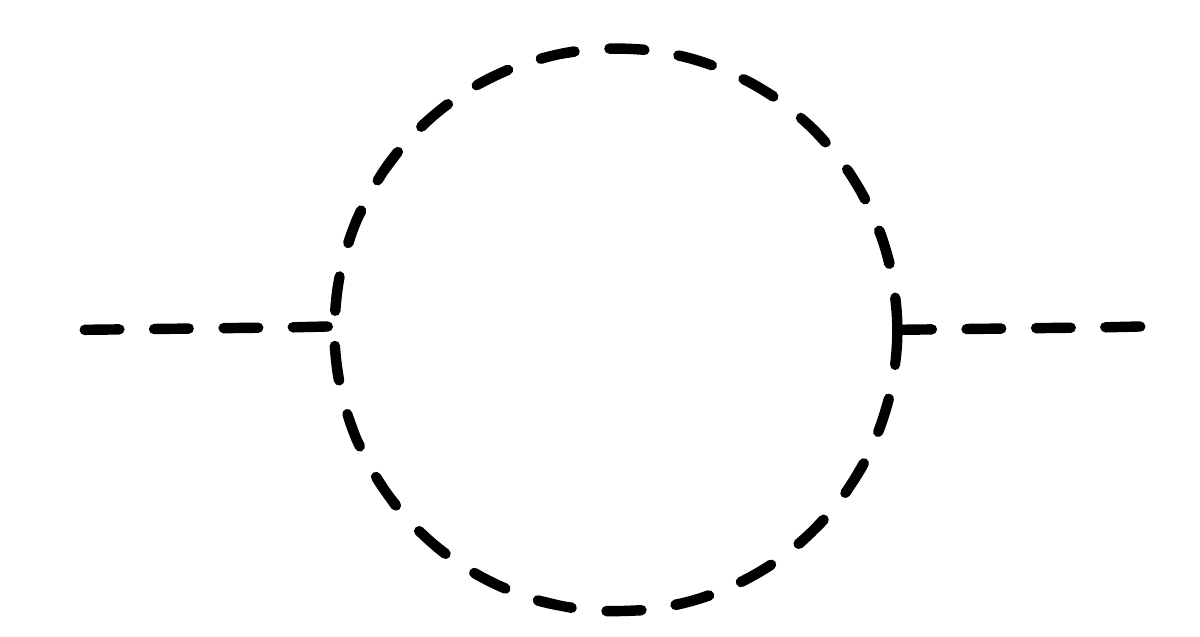}}
	\hspace{1cm}
	\subfloat[]{
		\label{LSMc}
		\includegraphics[width=0.2\textwidth]{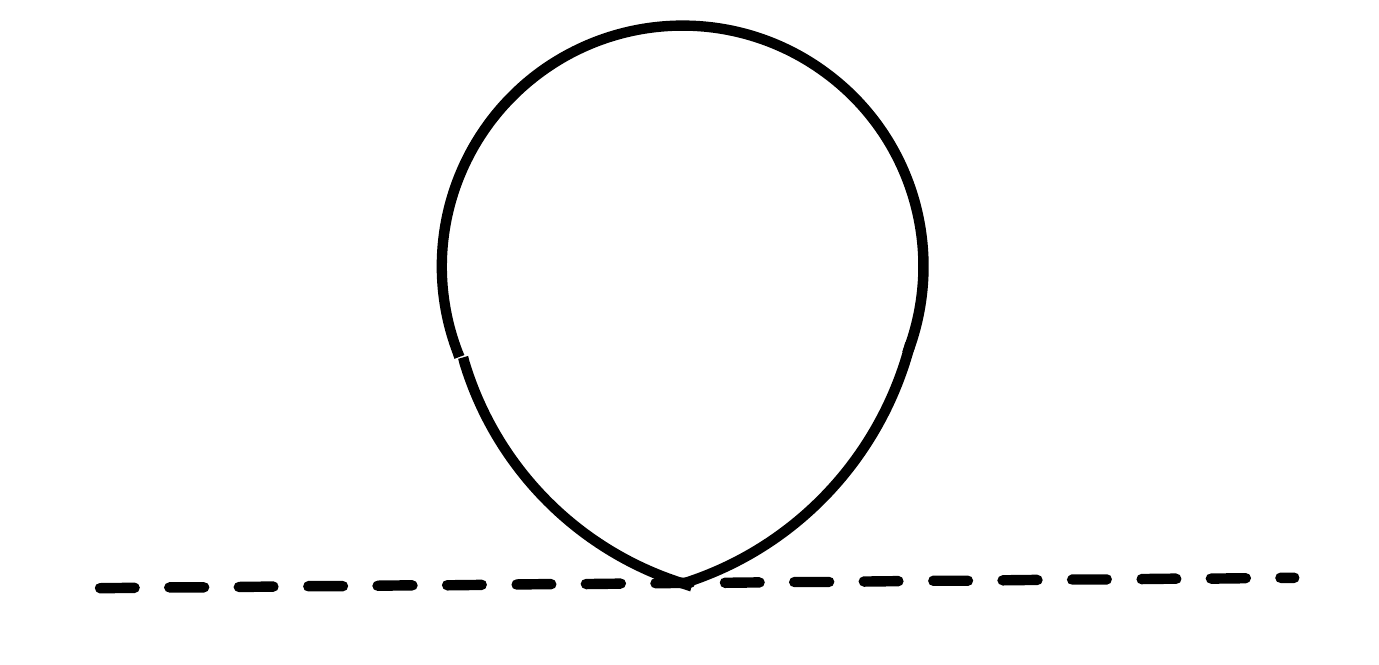}}
	\hspace{1cm}
	\subfloat[]{
		\label{LSMd}
		\includegraphics[width=0.2\textwidth]{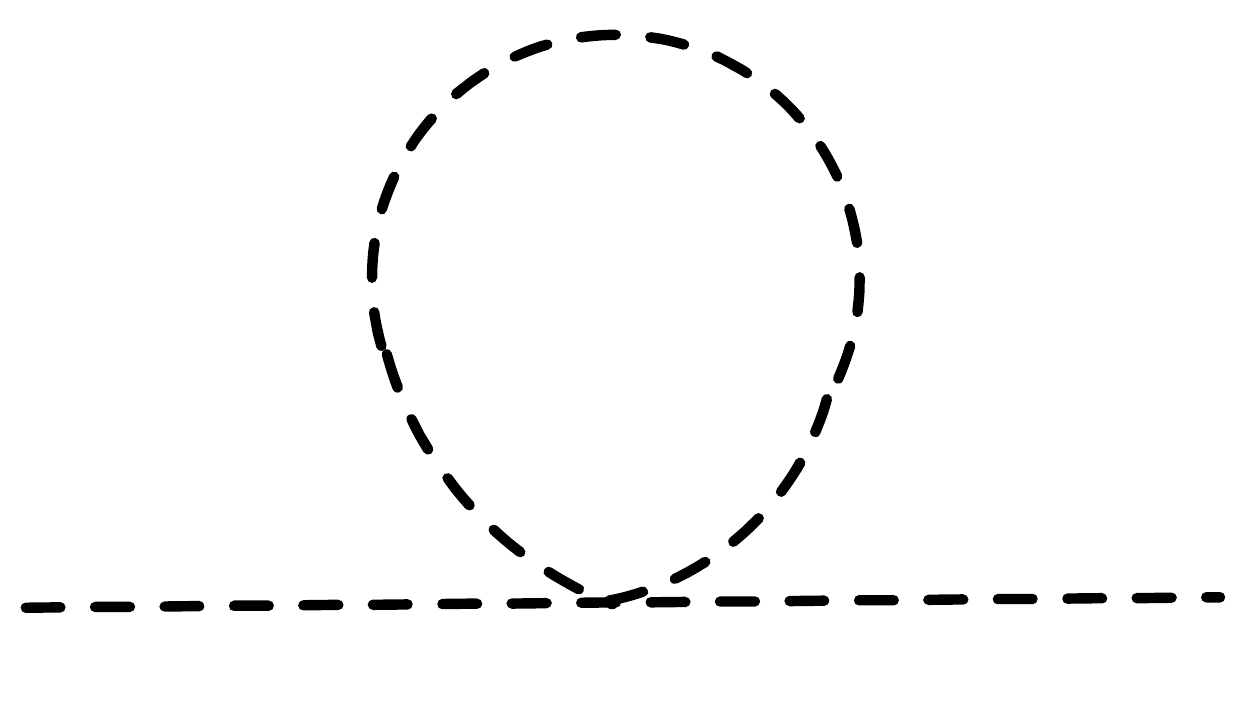}}
	\hspace{1cm}
	\subfloat[]{
		\label{LSMe}
		\includegraphics[width=0.2\textwidth]{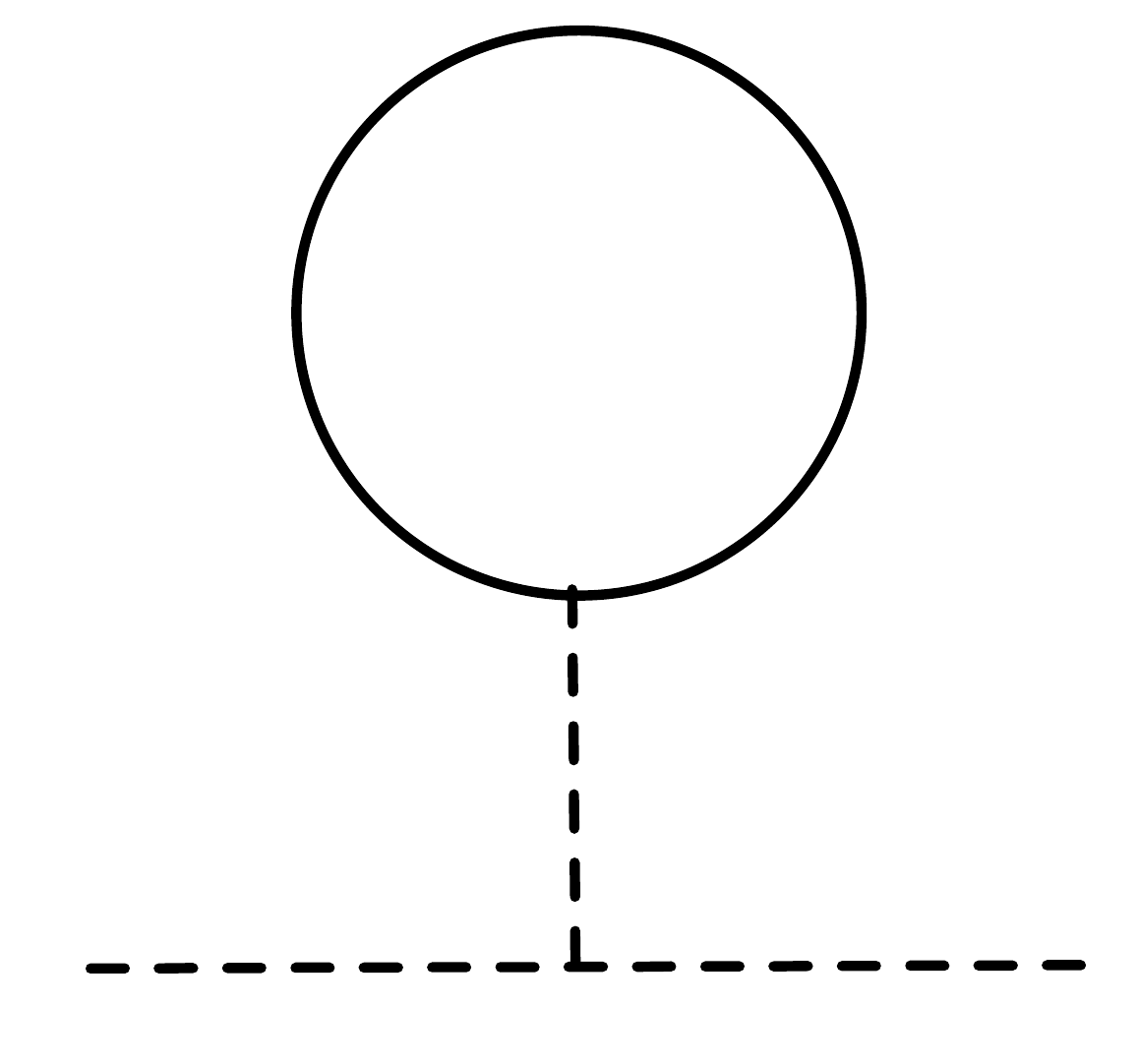}}
	\hspace{1cm}
	\subfloat[]{
		\label{LSMf}
		\includegraphics[width=0.2\textwidth]{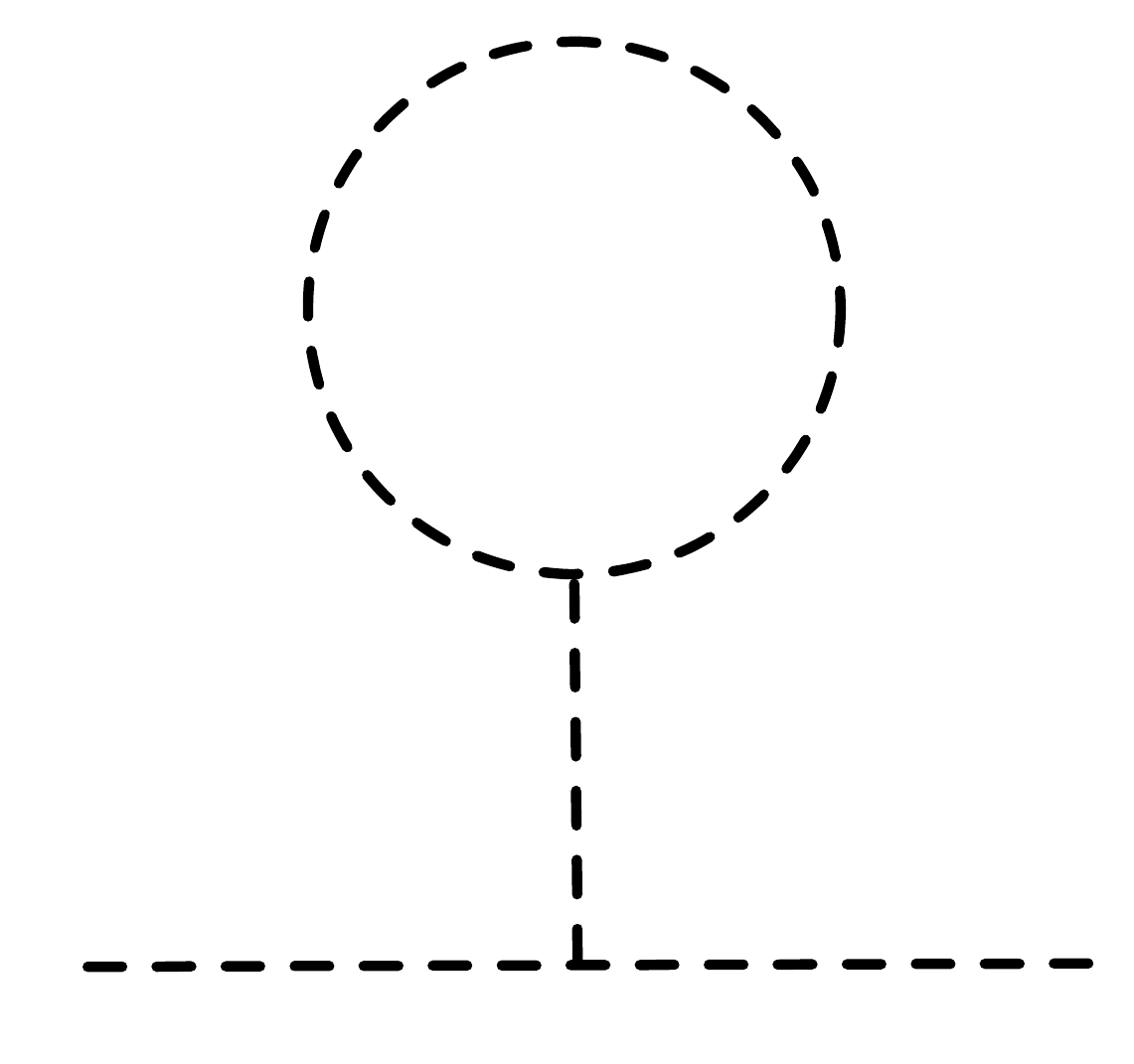}}
	\caption{One-loop contributions to the $\sigma$ self-energy in the LSM. The solid lines represent the pion fields and the dashed lines the sigma one.}
	\label{diagramLSM}
\end{figure}

\begin{equation}
\begin{split}
&\Sigma_a\left(k_0,\vec{k};T\right)=-3\lambda \left(M_{0\sigma}^2-M_{0\pi}^2\right) J\left(M_{0\pi};k_0,\vec{k},T\right),\\
&\Sigma_b\left(k_0,\vec{k};T\right)=-9\lambda  \left(M_{0\sigma}^2-M_{0\pi}^2\right)  J\left(M_{0\sigma};k_0,\vec{k},T\right),\\
&\Sigma_c\left(T\right)=3\lambda \hspace{0.1cm}G\left(M_{0\pi},T\right),\\
&\Sigma_d\left(T\right)=3\lambda \hspace{0.1cm}G\left(M_{0\sigma},T\right),\\
&\Sigma_e\left(T\right)=-9\lambda\dfrac{M_{0\sigma}^2-M_{0\pi}^2}{M_{0\sigma}^2}G\left(M_{0\pi},T\right),\\
&\Sigma_f\left(T\right)=-9\lambda\dfrac{M_{0\sigma}^2-M_{0\pi}^2}{M_{0\sigma}^2}G\left(M_{0\sigma},T\right).
\end{split}
\label{sigmaconts}
\end{equation}
where $J$ and $G$ are the finite-$T$ integrals of the bubble  and  tadpole diagrams respectively:
\begin{equation}
J\left(M_{i};k_0,\vec{k},T\right)=T\sum_{n=-\infty}^{\infty}\int\dfrac{d^3\vec{p}}{\left(2\pi\right)^3}\dfrac{1}{p^2-M_i^2}\dfrac{1}{\left(p-k\right)^2-M_{i}^2},
\label{Jter}
\end{equation}
\begin{equation}
G\left(M_i,T\right)=T\sum_n\int \dfrac{d^3\vec{p}}{(2\pi)^3}\dfrac{1}{\omega_n^2+\vec{p}^2+M_i^2},
\label{Gter}
\end{equation}
and where $\omega_n=2\pi n T$, $p=(i\omega_n,\vec{p})$, $k=(i\omega_m,\vec{q})$ and $i\omega_m\rightarrow k_0$ by analytic continuation, which can be performed after the Matsubara sums $\sum_n$ are carried out \cite{galekapustabook}. The divergent parts of the above loop integrals will be parametrized in DR. Their explicit expressions including  finite parts can be found in \cite{Gasser:1983yg} at $T=0$. At $T\neq 0$ we write $G\left(M,T\right)=G\left(M,T=0\right)+g_1(M,T)$ with the $g_1$ function defined in \cite{Gerber:1988tt}. As for the $J$ function, we will only need    $J\left(M;k_0,\vec{k}=\vec{0},T\right)$, whose finite temperature part   can be found  for instance in \cite{GomezNicola:2002tn,Nicola:2014eda} and for $k_0=0$ satisfies  $J\left(M;k=0,T\right) \equiv G_2(M,T)=-dG(M,T)/dM^2=G_2(M,T=0) +g_2(M,T)$ with $g_2(M,T)=-dg_1(M,T)/dM^2$.

As we are about to see, a standard LSM renormalization will allow us to express the self-energy as a finite quantity. Before that, we will provide a check of our result \eqref{susmod}.  From the lagrangian \eqref{lsm2} we have, within the perturbative $\lambda$ expansion, 

\begin{eqnarray}
v(T)&=&\frac{d}{dh}\log Z=-\frac{1}{2}\frac{d M_{0\sigma}^2}{dh} G(M_{0\sigma},T)-\frac{3}{2}\frac{d M_{0\pi}^2}{dh} G(M_{0\pi},T)-\frac{1}{4\lambda}\frac{d M_{0\pi}^4}{dh}+\frac{dv^2}{dh}M_{0\pi}^2
+v^2\frac{dM_{0\pi}^2}{dh}+\Od(\lambda^2)\nonumber\\
&=&v\left[1-\frac{3\lambda}{M_{0\sigma}^2}\left(G(M_{0\sigma},T)+G(M_{0\pi},T)\right)+\Od(\lambda^2)\right]\label{vTpertlsm}\\
&\Rightarrow& \condl(T)=-2B_0v^2 \frac{M_{0\sigma}^2}{M_{0\sigma}^2-M_{0\pi}^2} \left[ 1-\frac{3\lambda}{M_{0\sigma}^2}\left(G(M_{0\sigma},T)+G(M_{0\pi},T)\right)+\Od(\lambda^2)\right].
\label{condpertlsm}
\end{eqnarray}

The above result is actually compatible with the mean field approximation, where interactions and fluctuations are considered small (see \cite{Bochkarev:1995gi} in the chiral limit). Taking one more mass derivative, we obtain then the purely perturbative expression for the scalar susceptibility:

\begin{eqnarray}
\chi_S(T)=\frac{6B_0^2}{\lambda}\left[1+\lambda\left(3G_2\left(M_{0\sigma},T\right)+G_2\left(M_{0\pi},T\right)\right)+\Od(\lambda^2)\right].
\label{suspertlsm}
\end{eqnarray}

One can now check that we arrive exactly to the same result by expanding \eqref{susmod} in powers of $\lambda$ and using our previous results \eqref{vTpertlsm} and \eqref{sigmaconts}.

Our next step will be to provide a finite and scale-independent result for the self-energy at finite temperature. The DR pole can be absorbed in the $T=0$ renormalization of the  pion and sigma masses: 

\begin{equation}
M_{0\sigma}^2-M_{0\pi}^2=(M_\sigma^2-M_{\pi}^2)\left[1+\dfrac{6\lambda}{16\pi^2}\left(N_\epsilon+1-\log \frac{M_{\sigma}^2}{\mu^2}-\dfrac{1}{6}\right)\right]+O(\lambda^2),
\end{equation}
\begin{equation}
M_{0\pi}^2=M_\pi^2\left\lbrace1-\dfrac{3\lambda}{16\pi^2}\left[(N_\epsilon+1)\left(1-\dfrac{3M_{\pi}^2}{M_{\sigma}^2}\right)+\left(\dfrac{3M_{\pi}^2}{M_{\sigma}^2}-2\right)\log\frac{M_{\pi}^2}{\mu^2}+\log \frac{M_{\sigma}^2}{\mu^2}\right]\right\rbrace+O(\lambda^2),
\end{equation}
with $N_\epsilon = 2 /\epsilon- \gamma + \log 4 \pi $ and $\mu$ the DR renormalization scale. The above renormalization coincides with that in the chiral limit provided by \cite{Gasser:1983yg}. See also the $T=0$ calculation of the $\sigma$ propagator in \cite{Manohar:2008tc}.  With the above renormalization, we get the LSM one-loop self-energy finite and scale-independent:

\begin{eqnarray}
\Delta_\sigma^{-1}&=&M_\sigma^2+\Sigma(k_0,\vec{k};T),\nonumber\\
\Sigma(s,T=0)&=&\dfrac{3\lambda}{16\pi^2}(M_{\sigma}^2-M_{\pi}^2)\left[\sigma_{\pi}(s)\log \left(\dfrac{\sigma_{\pi}(s)+1}{\sigma_{\pi}(s)-1}\right)+3\hspace{0.1cm}\sigma_{\sigma}(s)\log \left(\dfrac{\sigma_{\sigma}(s)+1}{\sigma_{\sigma}(s)-1}\right)+\log\left(\dfrac{M_{\pi}^2}{M_{\sigma}^2}\right)-\dfrac{13}{3}\right]+\Od(\lambda^2),
\nonumber\\
\Sigma(k_0,\vec{k};T)&=&\Sigma(s,T=0)+ 3\lambda\left\{\frac{3M_\pi^2-2M_\sigma^2}{M_\sigma^2}\left[g_1(M_\pi,T)+g_1(M_\sigma,T)\right]\right.
\nonumber\\
&-&\left.\left(M_\sigma^2-M_\pi^2\right)\left[\delta J(M_\pi;k_0,\vec{k},T)+3\delta J(M_\sigma;k_0,\vec{k},T)\right]\right\}+\Od(\lambda^2),
\label{selfenergylsm}
\end{eqnarray}
where
\begin{equation}
\sigma_i(s)=\sqrt{1-\frac{4M_i^2}{s}}
\label{phasespace}
\end{equation}
is the two-particle phase space and $\delta J(M;k,T)=J(M;k,T)-J(M;k,0)$.

\begin{figure}
	\centering
	\subfloat{
		\includegraphics[width=12cm]{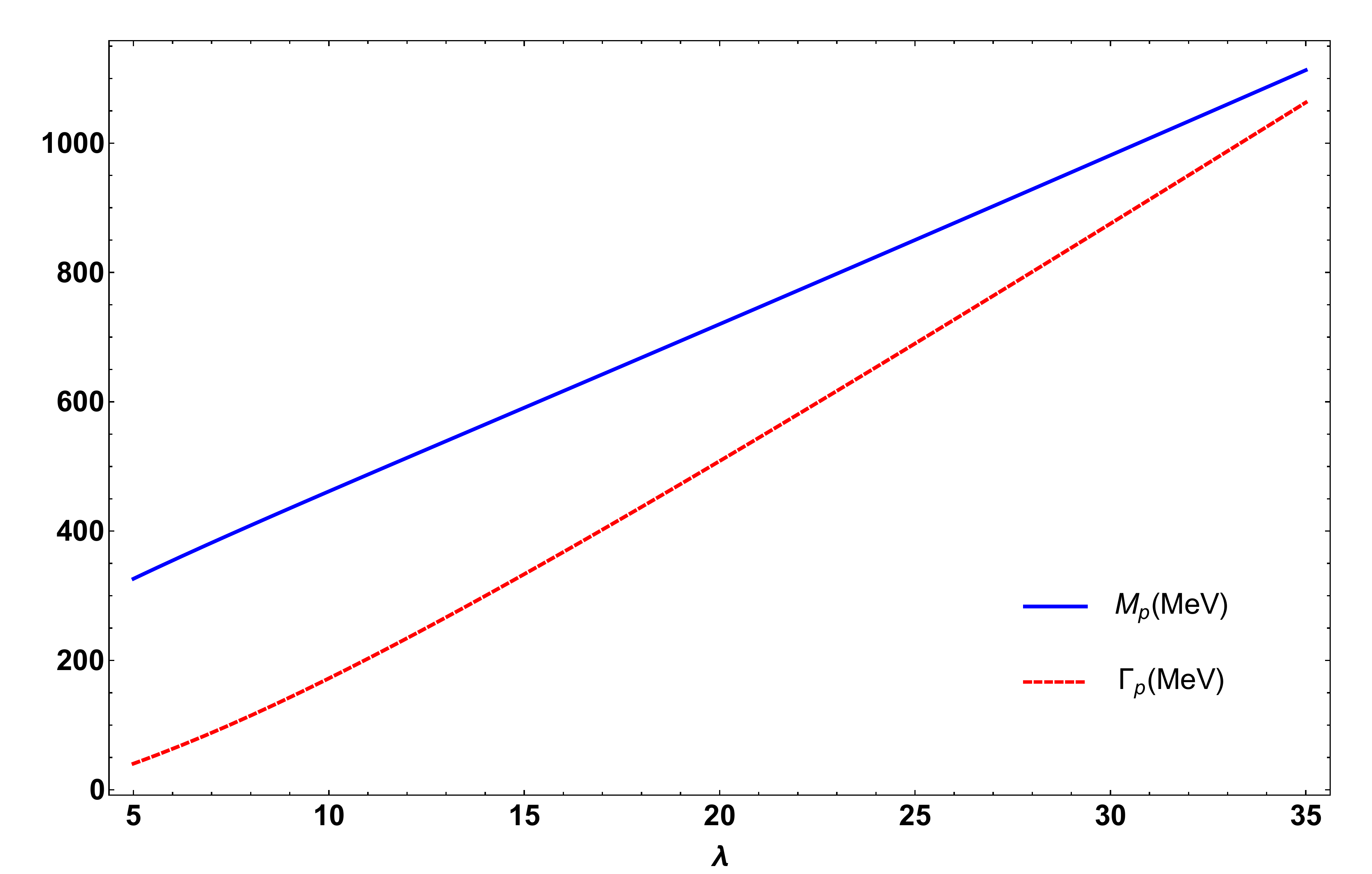}}
	\caption{Pole parameters $M_p$ and $\Gamma_p$ in the one-loop LSM as a function of $\lambda$ for the physical mass $M_\pi$.}
	\label{fig:mpgplsm}
\end{figure}

The pole of the propagator and its evolution with temperature can now be readily calculated. Perturbatively, the pole of the propagator is at $s_p=M_\sigma^2+\Sigma(k^2=M_\sigma^2)$, which we will parametrize as customary as $s_p=(M_p-i\Gamma_p/2)^2$. At $T\neq 0$, it will be enough for the purposes of this work to consider the pole at $\vec{k}=\vec{0}$. At $T=0$, we get, for $M_\sigma>2M_\pi$,

\begin{eqnarray}
\re{s_p}&=&  M_\sigma^2 + \frac{3\lambda\left(M_\sigma^2-M_\pi^2\right)}{16\pi^2}\left[-\frac{13}{3}+\sqrt{3}\pi+\log\left(\frac{M_\pi^2}{M_\sigma^2}\right)+\sigma_{\pi}(M_\sigma^2)\log \left\vert\dfrac{\sigma_{\pi}(M_\sigma^2)+1}{\sigma_{\pi}(M_\sigma^2)-1}\right\vert\right],\label{repolezeroTlsm}\\
\im{s_p}&=& -\frac{3\lambda\left(M_\sigma^2-M_\pi^2\right)}{16\pi}\sigma_{\pi}(M_\sigma^2),
\label{impolezeroTlsm}
\end{eqnarray}
which agrees with the result  in \cite{Masjuan:2008cp}  in the chiral limit. Following \cite{Pelaez:2015qba}, we will set our reference values for the numerical parameters of the model as those for which the $\sigma$ pole values lie near  the experimental determination for the $f_0(500)$ in the PDG \cite{Tanabashi:2018oca}, namely  $M_p^{PDG}\simeq (400-550)$ MeV, $\Gamma_p^{PDG}\simeq (400-700)$ MeV. As noted in \cite{Pelaez:2015qba}, there is no way to accomodate the LSM parameters to get good agreement both for $M_p$ and $\Gamma_p$. We show in Fig.\ref{fig:mpgplsm} the dependence on $\lambda$ of $M_p$ and $\Gamma_p$ for the physical $M_\pi=$ 140 MeV, which confirms the previous statement. In our numerical results  we have taken for the tree level $M_\sigma^2=M_\pi^2+2\lambda F_\pi^2$ with $F_\pi\simeq$ 93 MeV. In view of those results we select as a reference range for our numerical results the interval $\lambda\sim 10-20$ where the deviations from the PDG value are not large in either $M_p$ or $\Gamma_p$, the lower (higher) value of $\lambda$ favoring   $M_p$ ($\Gamma_p$) as showed for some sample values in Table \ref{masswidth}, where we also include the chiral limit values.   In turn, we note that, even though the typical $\lambda$ values needed are large, the one-loop corrections remain reasonably under control, lying  between 10-15 \% for the corrections to $M_p$ compared to the tree level $M_\sigma$.

\begin{table}[h!]
		\centering
		\begin{tabular}{ |c|c|c|c|  }
			\hline
			\hline
			$M_{\pi}$ (MeV)&$M_p$ (MeV)& $\Gamma_p$ (MeV) & $\lambda$\\
			\hline
			$0$ & $450.0$ & $172.5$ & $8.4$\\
			\hline
			$0$ & $775.1$   & $550.0$  & $20.0$\\
			\hline
			\hline
			\hline
			$140$ &$450.0$&$159.2$&$9.6$\\
			\hline
			$140$ & $750.1$ &$550.0$& $21.2$\\
			\hline
		\end{tabular}
		\caption{Pole mass and width at $T=0$ for sample values of $\lambda$, with $M_p=$Re $\sqrt{s_p}$ and $\Gamma_p=-2$Im $\sqrt{s_p}$.}
		\label{masswidth}
	\end{table}

Our next step will be to provide the results for the different approaches to the scalar susceptibility mentioned above.  Taking the $k\rightarrow 0^+$ limit for the self-energy, we can get $\chi_S(T)$ from the saturated approach \eqref{susgreen}. From  \eqref{selfenergylsm}, taking into account that 

$$\lim_{s\rightarrow 0}\sigma_i(s)\log\frac{\sigma_i(s)-1}{\sigma_i(s)+1}=-2,$$

we have:

\begin{eqnarray}
\Sigma(k=0;T)&=&\dfrac{\lambda}{16\pi^2}(M_{\sigma}^2-M_{\pi}^2)\left[11+3\log\left(\dfrac{M_{\pi}^2}{M_{\sigma}^2}\right)\right] +3\lambda\left\{\frac{3M_\pi^2-2M_\sigma^2}{M_\sigma^2}\left[g_1(M_\pi,T)+g_1(M_\sigma,T)\right]\right.
\nonumber\\
&-&\left.\left(M_\sigma^2-M_\pi^2\right)\left[g_2(M_\pi,T)+3g_2(M_\sigma,T)\right]\right\}+\Od(\lambda^2).
\end{eqnarray}

\begin{figure}
	\centering
	\subfloat{
		\includegraphics[width=9cm]{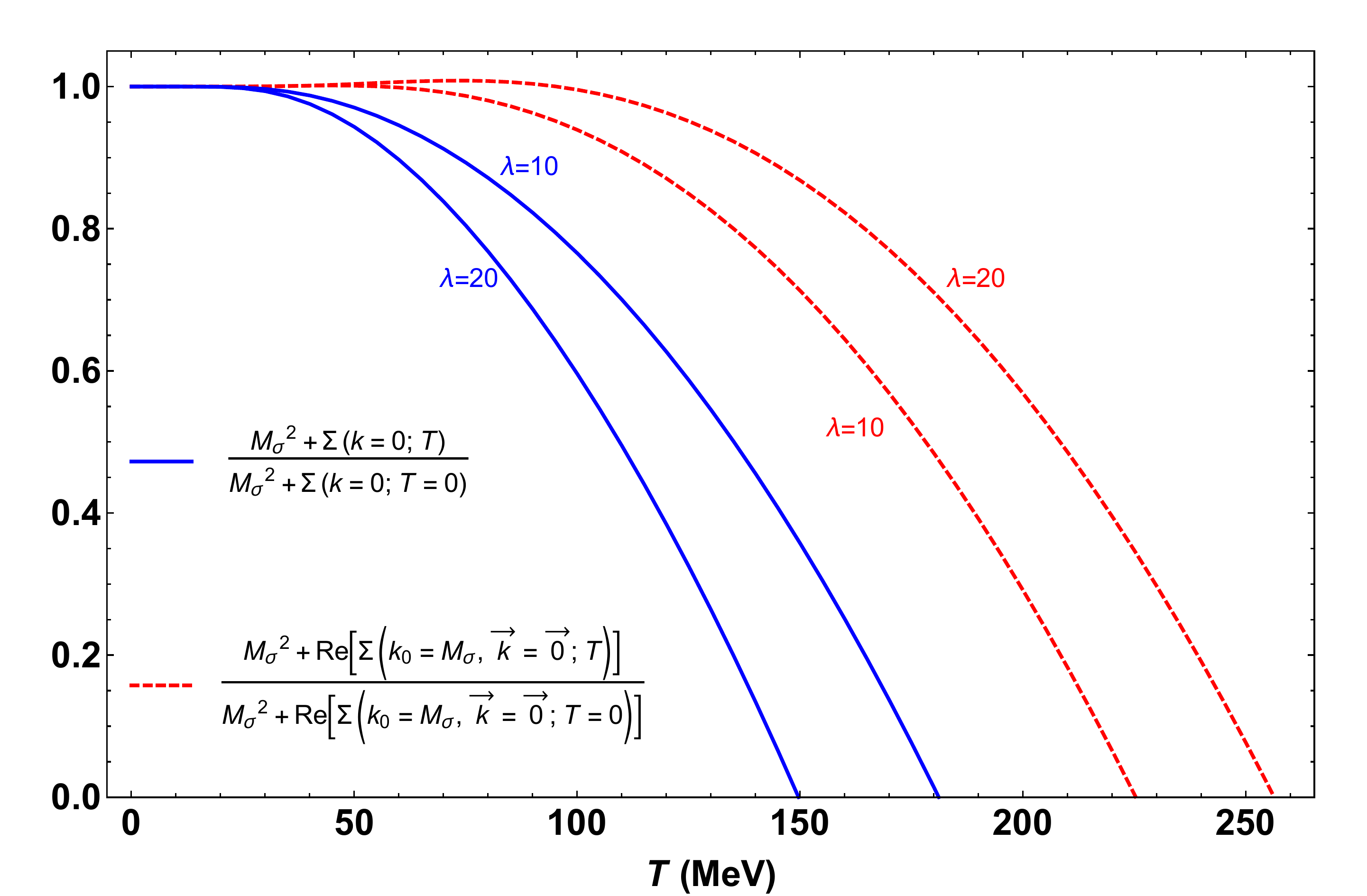}}
	\subfloat{
		\includegraphics[width=9cm]{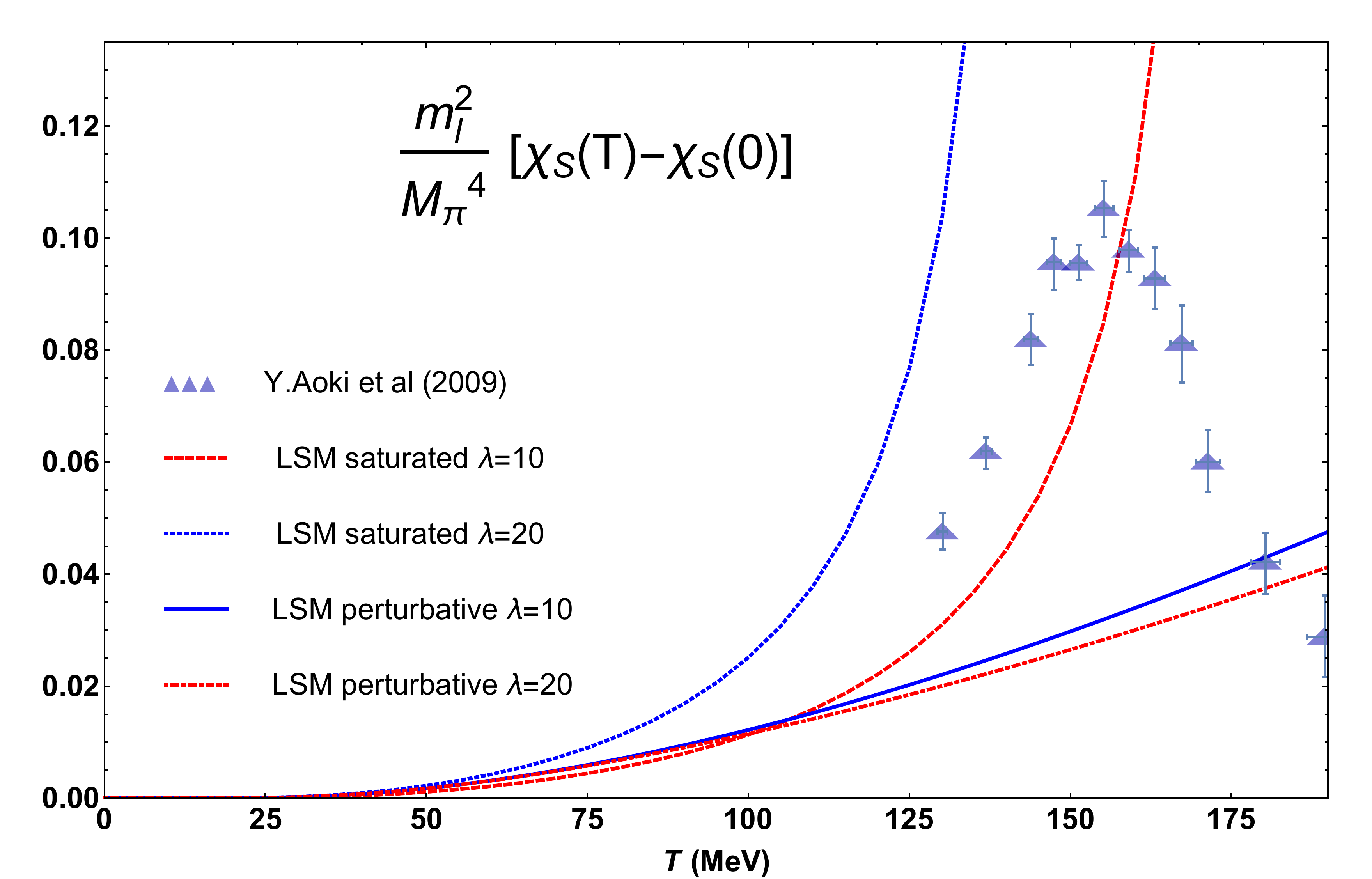}}
	\caption{Left: LSM thermal self-energy at $k=0$ and at $k_0=M_\sigma$, $\vec{k}=\vec{0}$, for the range $\lambda=10-20$. Right: Saturated susceptibility in the LSM compared to lattice data and to the purely perturbative one.  In both figures $M_\pi=140$ MeV. The lattice data and errors are from \cite{Aoki:2009sc}. }
	\label{fig:suscomplsm}
\end{figure}

In Fig.\ref{fig:suscomplsm} we show our results for $\chi_S(T)$ in the saturated approach, compared to the perturbative one arising from \eqref{suspertlsm}. For an easier comparison with lattice data 
 and with our results in sections \ref{sec:uchpt} and \ref{sec:hrg}, we are using for the saturated LSM susceptibility the following normalization:

\begin{equation}
\chi_S^{sat,LSM}(T)=A\frac{M_\pi^4}{4m_l^2}\dfrac{M_{0\sigma}^2+\Sigma\left(k=0;T=0\right)}{M_{0\sigma}^2+\Sigma\left(k=0;T\right)}.
\label{susunit}
\end{equation} 

For the results in Fig.\ref{fig:suscomplsm} we have taken $A=A_{ChPT}\simeq 0.15$ as in section \ref{sec:uchpt}  (see comments below). Nevertheless, the corresponding value for the normalization constant $A$ from the normalization given in  \eqref{susgreen} would be around $A_{LSM}\approx 4F_\pi^2/M_S^2(0)\approx 0.07-0.17$ for the range of values showed in Table \ref{masswidth} and hence compatible with the ChPT value.

The main conclusion that we extract from the results showed in Fig.\ref{fig:suscomplsm} is that the saturated approach provides a much stronger growth with temperature than the purely perturbative one, actually covering lattice data below the transition for the range of values of $\lambda\sim 10-20$ corresponding to the $T=0$ poles in Table \ref{masswidth}.  However, the saturated susceptibility actually diverges around the transition point, even in the massive case and therefore is not able to reproduce either the crossover peak. As we will see in section \ref{sec:uchpt}, the UChPT approach will considerably improve this behaviour. Nevertheless, an important comment is that even within this LSM approach, which has its own limitations as we are discussing here,  a saturated description of the scalar susceptibility in terms of the thermal self-energy seems to be able to describe lattice data reasonably without adding additional degrees of freedom. That feature is shared also by the UChPT approach and is one of the main conclusions of this work. Finally, in Fig.\ref{fig:suscomplsm} we have showed also a comparison of the $s=0$ result for the self-energy temperature dependence with the same approach evaluating the real part of the self-energy at $s=s_p$, the latter being the point at which the equivalent to the self-energy is naturally evaluated within the UChPT approach (see details in section \ref{sec:uchpt}). The qualitative behaviour is the same, although numerically the temperature at which the self-energy vanishes (divergent susceptibility) moves to a higher value. In both cases, the dropping behaviour can be understood as a chiral symmetry restoration tendency, since $M_\sigma^2+\re\Sigma$  corresponds to the $T$-dependent scalar mass, which is meant drops below the transition approaching the pion mass \cite{Bochkarev:1995gi}.

\subsection{Unitarized Chiral Perturbation Theory: thermal $f_0(500)$ saturation approach}
\label{sec:uchpt}

The LSM description discussed in the previous section relies on the one-loop $\lambda$ expansion. However, as we have just seen, the typical numerical $\lambda$ values needed to reproduce meson observables are large, in particular  to reconcile both the real and imaginary part of the $f_0(500)$ pole at $T=0$ found in the one-loop LSM with the experimentally observed range for those quantities. In addition, the $\sigma$ stable state in the lagrangian formulation is not well justified physically. 

A well-established framework to generate the $f_0(500)$, without having to appeal to an explicit $\sigma$-field lagrangian realization, is UChPT. One starts from the ChPT series for the $\pi\pi$ scattering amplitude in a given channel, projected into partial waves  of well defined isospin $I$ and angular momentum $J$ \cite{Gasser:1983yg}, namely $t^{IJ}(s,T)\simeq t^{IJ}_2(s)+t^{IJ}_4(s,T)+\dots$.  The $t_4$ contribution contains one-loop diagrams from the second-order ChPT lagrangian as well as tree level terms from the fourth order one  proportional to  Low Energy Constants (LEC). The temperature corrections arise in loops and are therefore included from the $t_4$ contribution onwards \cite{GomezNicola:2002tn}.   The ChPT  series  ensures the model-independent low-energy behaviour and is unitary only perturbatively, i.e. $\im t_4 =\sigma_\pi \vert t_2 \vert^2$ for $s\geq 4M_\pi^2$ and so on, with $\sigma_\pi$  the two-pion phase space defined in \eqref{phasespace}. 

An exactly unitary amplitude can be constructed by several methods. The main method we will follow here is the $\Od(p^4)$ Inverse Amplitude Method (IAM), originally developed at $T=0$ in \cite{IAM} and extended to finite temperature in \cite{Dobado:2002xf}. In that  approach, exact unitarity and matching to the low-energy ChPT expansion are demanded, including the finite temperature corrections to the scattering amplitude in the center of momentum frame,  which implies the following modification of the phase space:

\begin{equation}
\sigma_T(s,T)=\sigma_\pi\left[1+2n_B(\sqrt{s}/2,T)\right],
\label{thpsp}
\end{equation}
with $n_B(x,t)=\left[\exp(x/T)-1\right]^{-1}$ the Bose-Einstein distribution function, so that perturbatively $\im t_4(s,T) =\sigma_T (s,T) \vert t_2 (s) \vert^2$ for $s\geq 4M_\pi^2$.

Thus, the unitarized IAM partial waves read 

\begin{equation}
t_{IAM}(s;T)=\frac{t_2(s)^2}{t_2(s)-t_4(s,T)}.
\label{iam}
\end{equation}

The above  amplitude is analytic off the real axis and satisfies the exact thermal unitarity relation $\im t_{IAM}=\sigma_T \vert t_{IAM} \vert^2$ for $s\geq 4M_\pi^2$. As a consistency check, this relation has been  shown to hold exactly  within the large-$N_{GB}$ approach in the chiral limit, where $N_{GB}$ is the number of Goldstone Bosons \cite{Cortes:2015emo}. In addition, the IAM amplitude reproduces the ChPT series up to $\Od(p^4)$ when expanded at low energies and is  analytical in the complex $s$ plane \cite{Dobado:2002xf}, which  ultimately allows  to search for resonances as poles in the 2RS. Thus, the $f_0(500)$ ($I=J=0$) and the $\rho(770)$ ($I=J=1$) are generated at $T=0$ with their pole position parameters $s_p=(M_p-i\Gamma_p/2)^2$ in agreement with those quoted experimentally by the PDG \cite{Tanabashi:2018oca}. For the $f_0(500)$, taking the LEC given in \cite{Hanhart:2008mx}, which we will use throughout this work, one gets $M_p=442.66$ MeV and $\Gamma_p=433.0$ MeV at $T=0$.

According to our  discussion in section \ref{sec:lsm}, we expect the scalar susceptibility $\chi_S (T)$ to be proportional to the inverse of $\Sigma (k=0)$, $\Sigma$ denoting generically the self-energy of the $f_0(500)$ state. However, within the UChPT approach, the $f_0(500)$ state is dynamically generated and then emerges as a 2RS pole of the scattering amplitude rather than a time-ordered product or thermal correlator, as in the case of the LSM discussed in the previous section. Thus, within UChPT, instead of $\Sigma$ we  have access to the pole parameters of the $f_0(500)$ state, namely, $M_p$, $\Gamma_p$ and $g_{\sigma\pi\pi}$,  the effective $\sigma\pi\pi$ effective coupling \cite{Pelaez:2015qba}, so that the 2RS amplitude reads around the pole

\begin{equation}
t^{II}=\frac{1}{16\pi}\frac{g_{\sigma\pi\pi}^2}{s-s_p}+\dots,
\label{unitpoleexp}
\end{equation}
and the dots denote subdominant terms around $s\sim s_p$. Note that if we regard \eqref{unitpoleexp} as the exchange of a scalar state $f_0$, the self-energy of such state would satisfy  $\Sigma_{f_0}(s_p)=s_p$, where we have included in $\Sigma_{f_0}$ the equivalent of the tree-level mass. On the other hand, $\im\Sigma_{f_0} (k=0)=0$ since at $k=0$ there are no decay channels open, so that assuming that the sensitivity of $\re\Sigma_{f_0}$ from $s_p$ to $s=0$ lies within the typical uncertainty range of this approach which we will analyze in detail below, we are led to the following definition of the unitarized scalar susceptibility, which corresponds to the saturated thermal $f_0(500)$ state approach within UChPT:

 \begin{equation}
\chi_S^U(T)=A\frac{M_\pi^4}{4m_l^2}\frac{M_S^2(0)}{M_S^2(T)},
\label{susunit}
\end{equation}
where we follow the same normalization as in section \ref{sec:lsm}  and where the scalar thermal pole mass (defined as the real part of self-energy at the pole) is 

\begin{equation}
M_S^2(T)=\re s_p (T)=M_p^2(T)-\frac{1}{4}\Gamma_p^2(T),
\label{scalarmass}
\end{equation}
the temperature dependence of $M_p(T)$ and $\Gamma_p(T)$ being determined from the 2RS pole of the unitarized amplitude \eqref{iam}, as discussed above.    The  thermal mass definition \eqref{scalarmass} shows a dropping behaviour compatible with the expected chiral restoring features discussed in section \ref{sec:lsm}, unlike the $I=J=1$ channel, where the mass has a much softer $T$ dependence  \cite{Dobado:2002xf,Nicola:2013vma}. Moreover, in \cite{Nicola:2013vma}, it has been shown that if the normalization $A$ is chosen to match the perturbative ChPT one-loop result for $\chi_S$ at $T=0$, i.e., 

\begin{equation}A_{ChPT}=\frac{4m_l^2}{M_\pi^4}\chi^{ChPT}_S (0)=\frac{\chi^{ChPT}_S (0)}{B_0^2}\simeq 0.15,\label{Achpt}\end{equation}
the resulting $\chi_S^U$ follows closely the ChPT curve for low temperatures and develops a maximum at a temperature around $157$ MeV (with the LEC used in \cite{Nicola:2013vma}), supporting strongly our previous assumptions.  One of our main purposes here is to test in a  more quantitative way the reliability of that saturated  approach to describe lattice data, as compared with other approaches such as the HRG discussed in section \ref{sec:hrg}, the LSM described in  section \ref{sec:lsm}  or $\chi_S(T)$ obtained perturbatively in ChPT or the virial approach \cite{GomezNicola:2012uc}.

The theoretical uncertainties involved in $\chi_S^U$ in \eqref{susunit} can be parametrized  into three main types: the normalization factor $A$,   the choice of the unitarization method and the numerical uncertainties of the LEC involved in $\pi\pi$ scattering for the pole determination. Here we will analyze in detail the sensibility of this approach to those three sources, focusing on its description of lattice data at finite temperature while complying with the $T=0$ predictions for scattering data, the $f_0(500)$ pole and the ChPT low-energy approach.

Let us consider first the LEC dependence. As stated above, we will use as a reference set of LEC, those given in \cite{Hanhart:2008mx}, namely
\begin{equation}
l_1^r=-(3.7\pm 0.2) \times 10^{-3}, \quad l_2^r=(5.0\pm 0.4) \times 10^{-3}, \quad l_3^r=(0.8\pm 3.8) \times 10^{-3}, \quad l_4^r=(6.2\pm 5.7)\times 10^{-3},
\label{lecs}
\end{equation}
where $l_i^r$ are the SU(2) renormalized LEC according to the notation in \cite{Gasser:1983yg}, evaluated at a DR scale $\mu=770$ MeV. The above LEC were obtained by a fit of the IAM to scattering data, leaving $l_3^r$ and $l_4^r$ fixed to their original ChPT values in \cite{Gasser:1983yg}. We remark that the LEC appearing in the $\pi\pi$ scattering vertices are $l_{1,2}^r$, and are those to which the pole position parameters of resonances are most sensitive,  while $l_{3,4}^r$ arise from the renormalization of the pion mass $M_\pi$ and the pion decay constant $F_\pi$.

\begin{figure}
\centerline{\includegraphics[width=14cm]{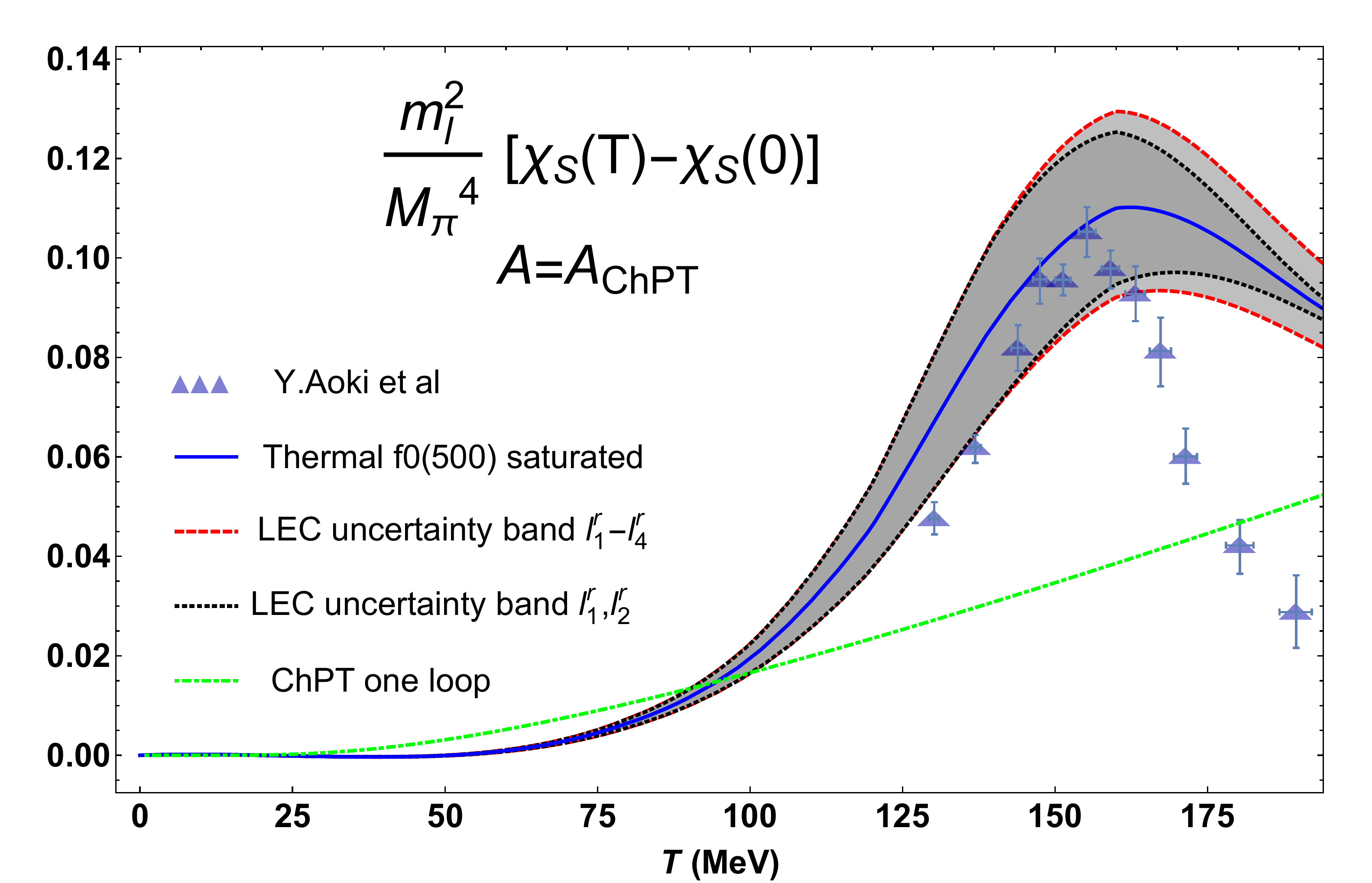}}
\caption{Results for the $f_0(500)$ saturated scalar susceptibility \eqref{susunit} normalized with $A_{ChPT}$ in \eqref{Achpt} including the uncertainties coming from the LEC \eqref{lecs}. The lattice data and errors are from \cite{Aoki:2009sc}. The central curves corresponding to the two uncertainty bands showed lie on top of each other.}
\label{fig:unitsuslec}
\end{figure}

We  will estimate the range of variation of the saturated susceptibility by considering, for every $T$,  the mean square error of the results obtained for  the eight combinations of upper and lower values given by \eqref{lecs}. The resulting uncertainty band is showed in Fig.\ref{fig:unitsuslec}, where the central line correspond to the average value and where we compare our prediction for $\chi_S$ based on the saturated thermal $f_0(500)$ approach, with lattice data coming from the work \cite{Aoki:2009sc}, the results of which can be easily translated into our present normalization. We consider also in the figure the uncertainty band generated solely by $l_1^r$ and $l_2^r$, which as we see remains very close to the band of the four LEC, confirming our previous observation about the sensitivity of the pole parameters to the LEC. We also include in the figure the one-loop ChPT curve with the LEC given in \cite{GomezNicola:2012uc}. Note that the ChPT result lies close to the LSM perturbative one in Fig.\ref{fig:suscomplsm} since the leading behaviour in the LSM result \eqref{suspertlsm} comes from $g_2(M_\pi,T)$ which is precisely the ChPT pion gas contribution \cite{GomezNicola:2012uc}.

These results lead to  interesting conclusions:  as we had anticipated, the saturated UChPT result reproduces the expected crossover peak around the transition and hence improves  over the saturated LSM in section \ref{sec:lsm}. We will actually see here that the adequate description of the $T=0$ pole of the UChPT result, as well as other basic requirements such as analiticity and unitarity, produce this behaviour.  Moreover, most of the lattice data fall into the uncertainty band, the approach being especially adequate near the transition region. Put in different words, one could use the LEC as fit parameters to reproduce the lattice $\chi_S$ at finite temperature and the results of the fit would be in the range allowed by the $T=0$ determinations of those LEC based on experimental information.

Nevertheless, it would be reassuring to consider other ways to test the robustness of the unitarized approach. For that reason, let us consider  another possible theoretical source of uncertainty, the unitarization method, which in turn will allow us to understand better which are the essential requirements that we should incorporate in the unitarized approach.

As mentioned above, the IAM satisfies unitarity for partial waves and reproduces the two first terms of the ChPT series at low energies, i.e. $t_2+t_4$. If we relax the second condition only to $t_2$, this leads to the so called $K$-matrix amplitude (see for instance the discussion in \cite{Delgado:2015kxa,Pelaez:2015qba}):

\begin{equation}
t_K(s;T)=\frac{t_2(s)}{1-\sigma_T (s,T)t_2(s)}.
\label{Kmat}
\end{equation}
where we have used that $\im t_4(s,T)=\sigma_T t_2(s)^2$ for $s\geq 4M_\pi^2$. 
Although the above  amplitude is unitary, it is not analytic due to the phase space factor $\sigma_T(s,T)$, so it cannot be properly extended to the complex $s$-plane, in particular to define properly the 2RS.  The requirement of analyticity is then crucial, as satisfied for instance by the IAM. With some modifications with respect to \eqref{Kmat}, we can construct a unitary, analytical amplitude, different from the IAM, and based on the so called chiral unitary approach \cite{Oller:1997ti} as follows:

\begin{equation}
t_{U{mod}}(s;T)=\frac{t_2^2(s)}{t_2(s)-t_{4J}(s,T)}.
\label{tunitmod}
\end{equation}
with

\begin{equation}
t_{4J}(s,T)=t_4(s,0)+16\pi t_2(s)^2 \left[J(M_\pi;k_0=\sqrt{s},\vec{k}=\vec{0},T)-J(M_\pi;k_0=\sqrt{s},\vec{k}=\vec{0},T=0)\right],
\end{equation}
and  $J$ the loop thermal integral defined in \eqref{Jter}, which comes from   the $s$-channel $\pi\pi$ scattering amplitude in the center of momentum frame, responsible for the unitarity contribution \cite{GomezNicola:2002tn}. 

The unitarized amplitude \eqref{tunitmod} can  be understood as obtained from \eqref{Kmat} by replacing the $\sigma_T (s,T)$ contribution in the denominator by an analytic function in $s$ satisfying unitarity, since $\im J(s,T)=\sigma_T(s,T)/(16\pi)$ for $s\geq 4M_\pi^2$.  Note that we keep the full $t_4$ ChPT amplitude at $T=0$. The reason is that we are renormalizing the $T=0$ divergent part of the integral \eqref{Jter}, dimensionally regularized, following the standard ChPT prescription \cite{Gasser:1983yg}, i.e, absorbing the divergence in the LEC. On the other hand, since we are using the LEC in \cite{Hanhart:2008mx}, fitted with the full IAM, to be consistent we have to ensure that the modified amplitude \eqref{tunitmod} reduces at $T=0$ to the IAM one in \eqref{iam}. This guarantees also that the $T=0$ $f_0(500)$ pole remains at the same value, compatible with the PDG, with these two different unitarization methods at finite temperature. In addition, in this way we will be able to test again the sensitivity to the LEC uncertantities in \eqref{lecs}.  As for the finite temperature correction in \eqref{tunitmod}, we are taking the minimal contribution ensuring unitarity and analyticity, i.e., the $T$-dependent part of $J(s,T)$.  As we are about to see, keeping the three requirements of analiticity, unitarity and the $T=0$ pole lead to a qualitative behaviour compatible with the crossover.

\begin{figure}
\centerline{\includegraphics[width=9cm]{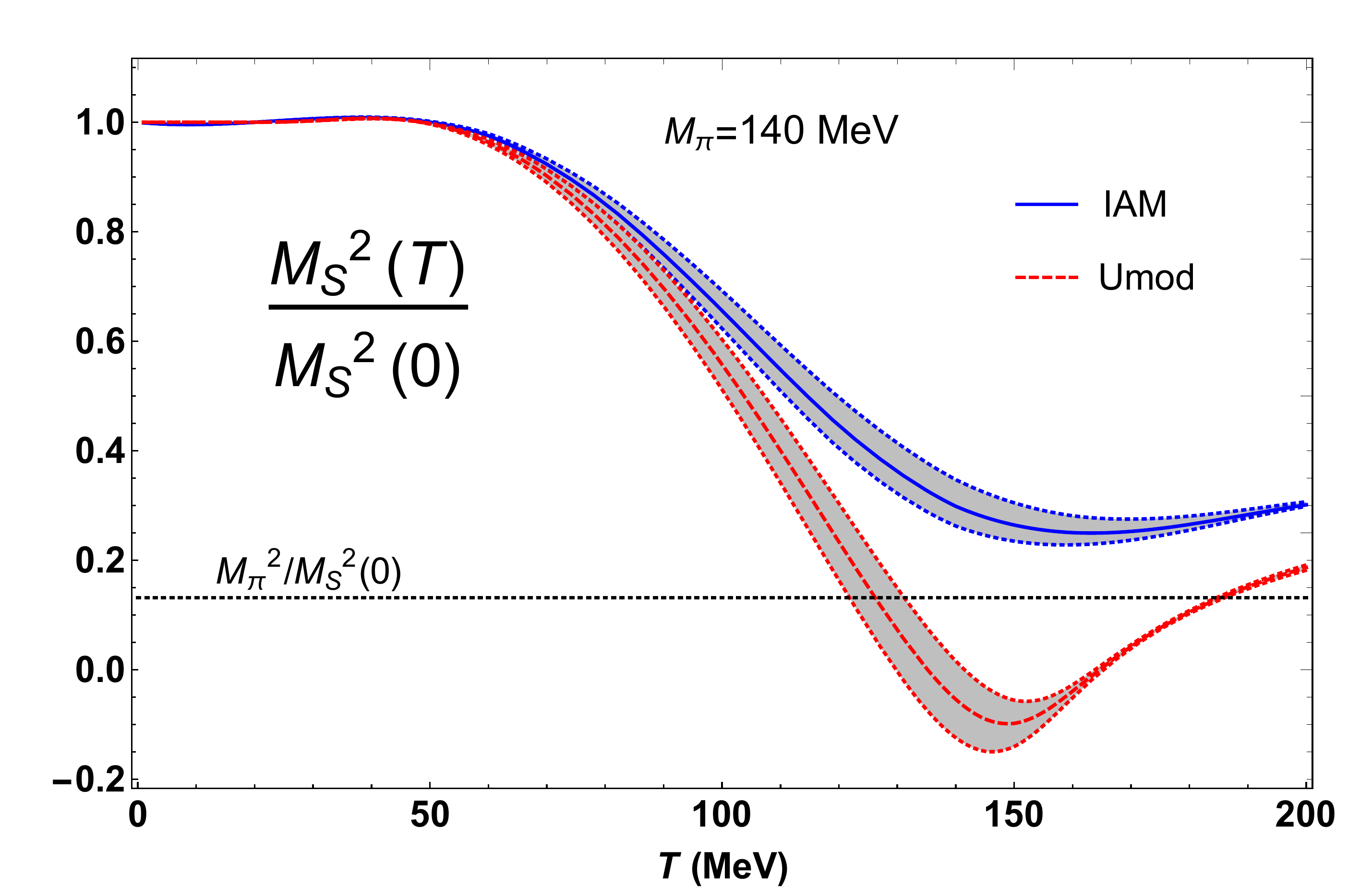}
\includegraphics[width=8.8cm]{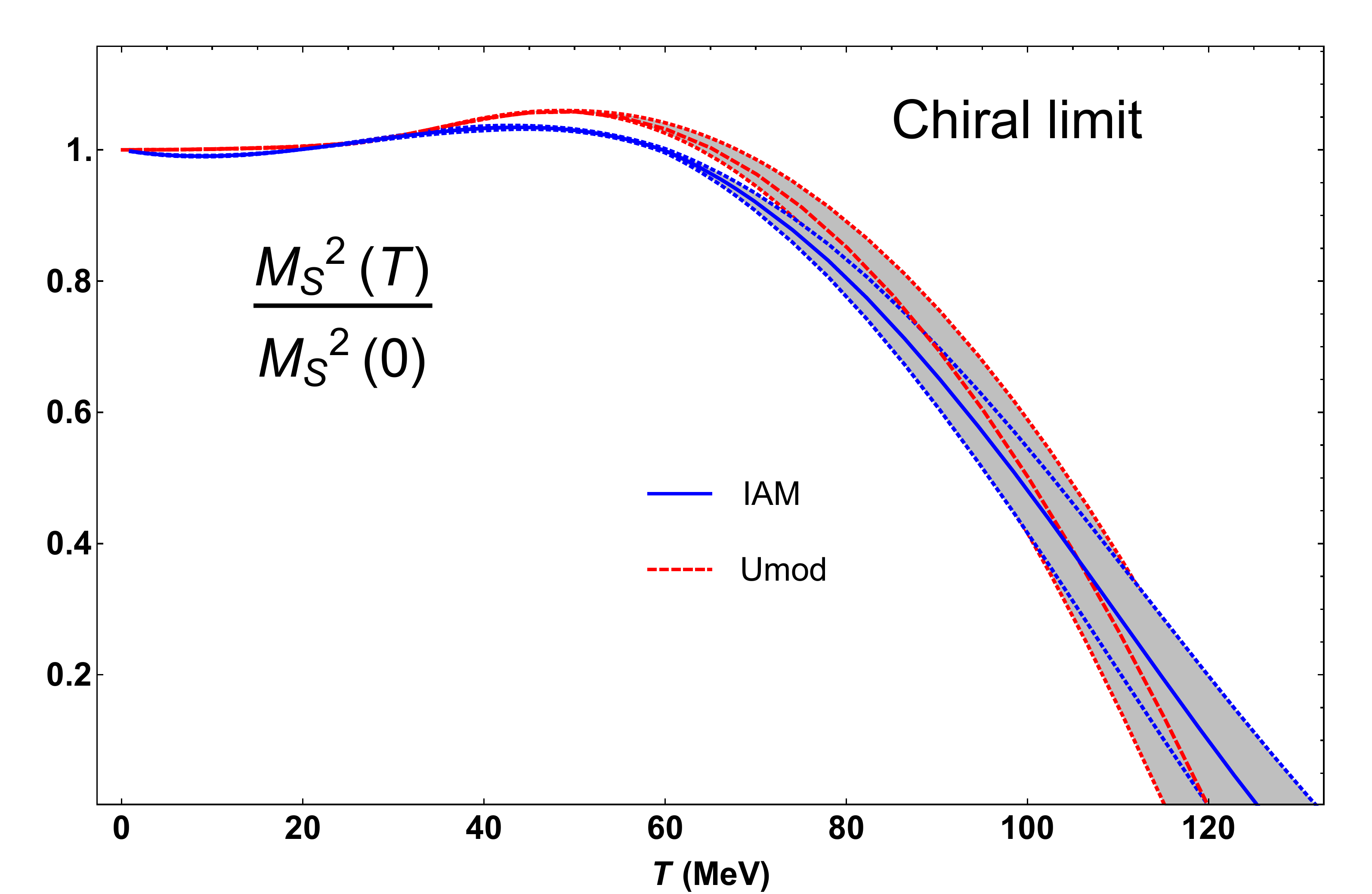}}
\caption{Squared scalar $f_0(500)$ thermal mass as defined by \eqref{scalarmass}, calculated with the  IAM and $U_{mod}$ unitarization methods in \eqref{iam}  and \eqref{tunitmod} respectively. 
Left: for the physical pion mass, where we show the pion mass squared value. Right: in the chiral limit. In both cases, the bands reflect the uncertainties in the LEC $l_1^r$ and $l_2^r$ given in \eqref{lecs} with $l_3^r$ and $l_4^r$ fixed at their central values.}
\label{fig:Msq}
\end{figure}

In Fig.\ref{fig:Msq} we plot the resulting $M_S^2(T)$ defined in \eqref{scalarmass}, with the two methods we have discussed, i.e., the IAM and the $U_{mod}$ ones. We also consider the same function taking the light chiral limit ($M_\pi\rightarrow 0^+$) in our expressions (recall that the $l_i^r$ in \eqref{lecs} are mass-independent). We observe that the qualitative behaviour around the transition s the same with both methods, i.e, they both develop a minimum around $T\simeq$ 150 MeV in the massive case. However, we see that the curve of the $U_{mod}$ method reaches zero before the minimum, which would give rise to a divergent susceptibility at that point, pretty much like the LSM saturated approach in section \ref{sec:lsm} except that in the LSM the thermal mass does not develop a minimum.  That difference between the two methods remains when the uncertainty bands for the LEC are included.  In that figure, we also show the constant $M_\pi^2$ reference value. The fact that the two methods give rise to a decreasing function approaching the pion mass squared strengthens the interpretation of $M_S^2(T)$ as a scalar mass, since $O(4)$ restoration would imply the degeneration of $\sigma-\pi$ states, while the thermal dependence of the pion mass is meant to be smooth \cite{Schenk:1993ru}.

Therefore, the  requirements of thermal unitarity and analyticity, together with the $T=0$ pole prediction, guarantee the key qualitative features of a crossover behaviour in terms of the  position of the minimum as compared to the lattice prediction for $T_c$, reasonably maintained within the LEC uncertainty band. This is  then a robust result. However, the additional requirement, only fulfilled by the IAM, of complying with the ChPT thermal scattering amplitude up to  fourth order is needed to describe  the scalar  susceptibility accurately. 

The previous difference between the two methods is washed out in the chiral limit, as Fig.\ref{fig:Msq}  shows. There, both methods yield results compatible among them within the LEC uncertainty band, predicting a critical temperature where the scalar mass vanishes, which is consistent with the expected reduction of the lattice $T_c$ of around $15-20 \%$ in the chiral limit \cite{Bazavov:2011nk}. 

Our results in this section might seem somehow striking, since we are describing a thermodynamical observable near the transition with just one effective state, the thermal $f_0(500)$, without using the information coming from the rest of the hadronic spectrum. Qualitatively, we had observed the same feature in the LSM in section \ref{sec:lsm}. Actually, this is our main motivation to compare the UChPT analysis with the the HRG approximation described in section \ref{sec:hrg}.  In this regard, one must take into account that, as stated above,  $\chi_S (T)$ is precisely the observable where the lightest scalar state is meant to dominate, while this conclusion may not be extensible to other quantities such as the quark condensate. Thus, the thermal unitarization procedure seems to incorporate in a natural way the relevant information of higher order states, encoded precisely in the LEC. Our results may look at first at odds with the usual claim within the HRG approach stating that the $f_0(500)$ state can be ignored in the list of hadron states contributing to the partition function,  motivated partly by a cancellation between the $IJ=00$ and $IJ=20$ channels in the $T=0$ partial waves  when considering the virial expansion \cite{GomezNicola:2012uc,Broniowski:2015oha} in which those channels appear weighted by the  $(2I+1)(2J+1)$ factor. However, it is important to point out that we are including here, as a key ingredient,  the {\em thermal} corrections to the $\pi\pi$ scattering amplitude, which include higher order finite-$T$ corrections not included in the usual virial approach, where scattering is included only at $T=0$. Those thermal corrections, as explained above, account for the thermal unitarity processes giving rise ultimately to the main modifications of the $f_0(500)$ pole parameters, directly connected with chiral symmetry restoration as we  are seeing here.

In connection with the last comment, one may wonder what would be the  effect of the $f_0(500)$ state and its thermal modifications in other observables relevant in Heavy Ion Collisions, such as hadron multiplicities and yields. The latter  have been successfully described within the so called thermal statistical models, very much in the same spirit as the HRG \cite{Andronic:2005yp,Andronic:2008gu,Floris:2014pta,Andronic:2017pug} (see section \ref{sec:hrg}) where the different PDG states contribute through their free partition function, and the resonances width can be incorporated by integrating in energy with a suitable Breit-Wigner shape \cite{Andronic:2005yp}. The decay channels of those resonances feed the hadron yields at chemical freeze-out.  Following this approach, the effect of including the $f_0(500)$ was first studied in \cite{Andronic:2008gu}, resulting in a few percent increase in the pion yield  from the $\pi\pi$ decay channel. That analysis showed also little dependence on variations of the $f_0(500)$ mass and width, which would lead to the conclusion that the finite-$T$ corrections we are discussing here would not have a significant effect for those observables. Regarding the connection with the QCD phase transition, hadron yields and multiplicities are correlated to hadronization rather than  to chiral symmetry  \cite{Andronic:2017pug} and hence we would expect a smaller effect of the  $f_0(500)$ modifications  addressed in the present work.  

However, as we have discussed above, the $f_0(500)$ should be treated as a broad resonance and hence the Breit-Wigner approach is not quite adequate in that case. In addition, as we have just commented, the $IJ=20$ repulsive channel would produce a cancellation of the  $f_0(500)$ effect that has to be accounted for. Such analysis has been performed in \cite{Broniowski:2015oha} within the virial approach, showing that such cancellation takes place also for the pion yield, resulting in a much smaller effect, around $0.3\%$ decrease. Within that approach, the quantity controlling the particle spectra for those channels and hence the pion yield and multiplicity is

\begin{equation}
d_{IJ}(E)=\frac{1}{\pi}\frac{d \delta_{IJ}(E)}{d E}
\label{distpion}
\end{equation}
with $E=\sqrt{s}$ and $\delta_{IJ}$ the corresponding $IJ$  channel phase shift.  Thus, in order to provide here a rough estimate of the possible effects of the $f_0(500)$ spectral modifications, we have calculated the $d_{00}(E)$ and $d_{20}(E)$ distributions in \eqref{distpion} with the phase shifts obtained from the IAM, both at $T=0$ and at $T=156$ MeV, which is the freeze-out temperature considered in  \cite{Broniowski:2015oha}. The result is that the qualitative picture that we have just described with $T=0$ phase shifts does not change much for the isospin-weighted combination $(2I+1)(2J+1)d_{IJ}$,  the modification being smaller as $E$ increases. The modifications for the individual $d_{IJ}$ follow a similar behaviour.

The previous arguments indicate  that the thermal effects on the $f_0(500)$ discussed here are not expected to produce large corrections regarding the pion multiplicity and yield, unlike  the case of the scalar susceptibility. A different story though would be the study of correlations such as $\pi^+\pi^-$, which are not isospin averaged so that   the previous cancellation does not occur \cite{Broniowski:2015oha}. In this sense, a promising line of research is the analysis of correlations and fluctuations in Heavy Ion Collisions and their connection with the QCD phase diagram \cite{Luo:2017faz}. 

Generally speaking, it would be interesting to examine how the thermal dependence of other PDG states can affect different observables, including hadron yields and correlations. Although such analysis is beyond the scope of this work,  there are significant examples of light mesons which might be of interest.  Apart from the $\rho$ meson and its well-known influence in the photon and dilepton spectrum, already mentioned in the introduction, other relevant states  which are  in the line of our present approach are the $\kappa$ and $a_0$ mesons which play a crucial role  to understand the pattern of chiral symmetry restoration in connection with the $U(1)_A$ symmetry \cite{GomezNicola:2017bhm}.

\section{Hadron Resonance Gas approach} 
\label{sec:hrg}

The results in the previous sections show that one can actually describe correctly the lattice results for the scalar susceptibility, saturating it with just the contribution from the thermal $f_0(500)$ state. On the other hand, one would expect that any thermodynamical quantity should be sensitive to higher order hadron states as the transition point is approached, according to the standard framework of the HRG, as we have mentioned before. Therefore, the inclusion of the thermal effects in the $f_0(500)$ pole and the LEC dependence somehow account effectively for the effect of those states, generating novel additional features such as the crossover-like behaviour discussed in the previous section. To make this comparison more clear, we will provide in this section the result for the scalar susceptibility within the HRG approach, which, as stated in the introduction, has been used extensively in the literature to describe the hadron gas below the transition.

The free energy density in the simplest HRG approximation, i.e., considering only free resonant states without including their width nor their interactions, is given by \cite{Karsch:2003zq,Karsch:2003vd,Tawfik:2005qh,Leupold:2006ih,Huovinen:2009yb,Jankowski:2012ms}

\begin{eqnarray}
z(T)&=&z_M(T)+z_B(T),\nonumber\\
z_{M,B}(T)&=&\pm T \sum_{M,B} d_i \pint \log\left[1\mp e^{-\beta E_i(p)}\right],
\label{zHRG}
\end{eqnarray}
where $E_i=\sqrt{\vert\vec{k}\vert^2+M_i^2}$, $M,B$ stand for the meson and baryon contributions, the upper sign is for mesons and the lower one for baryons. The sum extends to  hadron states with degeneracy $d_i$ and mass $M_i$ quoted in the PDG \cite{Tanabashi:2018oca}. In this work, we will consider only hadron states up to $M= 2$ GeV, following \cite{Jankowski:2012ms}.

From the pressure or the free energy density one can  in principle derive straightforwardly the quark condensate and the scalar susceptibility according to \eqref{condef}-\eqref{susdef}. However, the HRG formulation is parametrized in terms of hadron masses, so that any calculation involving quark mass derivatives requires modeling the hadron mass dependence on quark masses. Several approximations for such dependence have been followed in the literature within the HRG context, starting from a simple linear dependence of the form $\frac{\partial M_h}{\partial M_\pi^2}=2C$ with constant $C$ 
\cite{Karsch:2003zq,Tawfik:2005qh} to more elaborated ones \cite{Karsch:2003vd,Leupold:2006ih,Huovinen:2009yb,Jankowski:2012ms}. Here, we will follow the approach in \cite{Leupold:2006ih,Jankowski:2012ms}, which gives a good fit for the quark condensate to the $N_t=12$ lattice data in \cite{Aoki:2009sc}, used here as a lattice reference set of data both for the quark condensate and for the scalar susceptibility.  Within that approach, the dependence of pseudo Nambu-Goldstone Bosons, i.e, pion, kaon and eta masses, is extracted directly from the one-loop ChPT calculation \cite{Gasser:1984gg}, while the masses of the rest of hadrons are taken to scale within a constituent quark picture  as

\begin{eqnarray}
\frac{\partial M_B}{\partial m_{l,s}}&=&(3-N_s)\frac{\partial M_l}{\partial m_{l,s}}+N_s\frac{\partial M_s}{\partial m_{l,s}},\nonumber\\
\frac{\partial M_M}{\partial m_{l,s}}&=&(2-N_s)\frac{\partial M_l}{\partial m_{l,s}}+N_s\frac{\partial M_s}{\partial m_{l,s}},
\end{eqnarray}
where the constituent masses $M_l$, $M_s$ for light and strange quarks are extracted from the Nambu-Jona-Lasinio model calculation in \cite{Blaschke:2011yv}. We follow \cite{Jankowski:2012ms} for the assignments of the hadron strangeness content for open and hidden strange mesons, as well as for singlet and octet members. 

\begin{figure}
\centerline{\includegraphics[width=9cm]{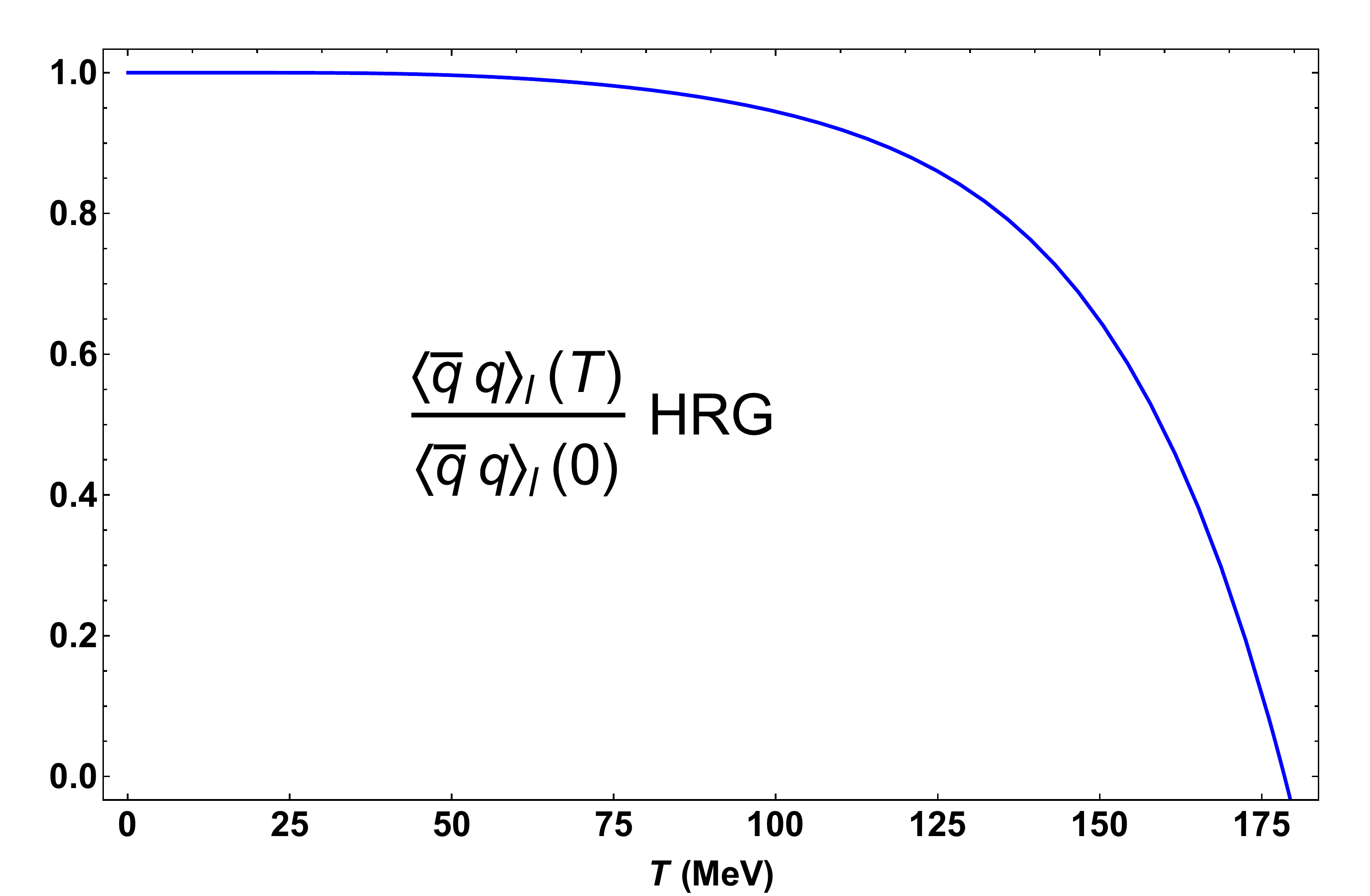}
\includegraphics[width=9cm]{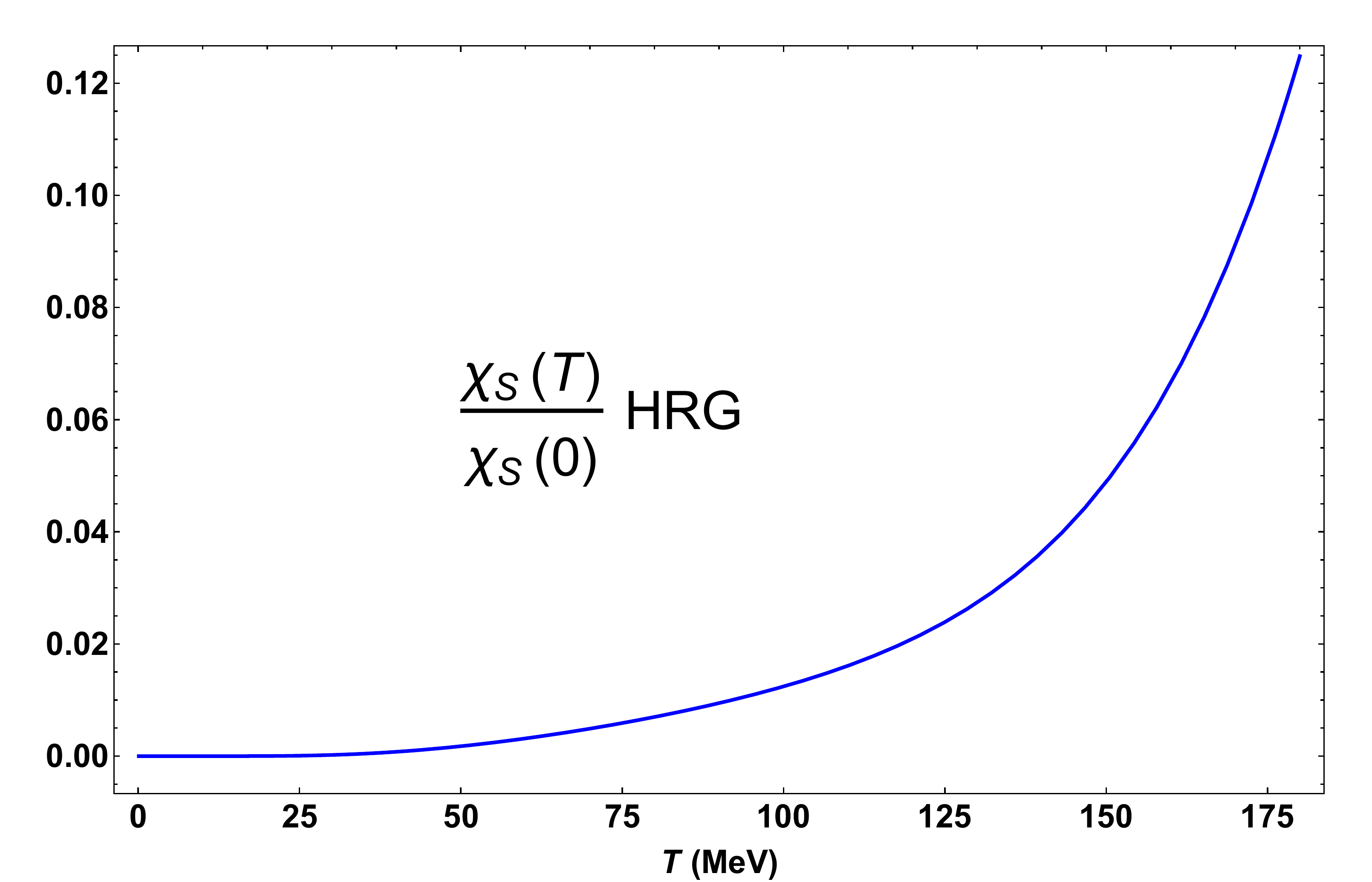}}
\caption{Light quark condensate and scalar susceptibility in the Hadron Resonance gas approach described in the main text.}
\label{fig:condsushrg}
\end{figure}

We show in Fig.\ref{fig:condsushrg}  the light quark condensate and the scalar susceptibility within the HRG approach. As it is known, within this approach the quark condensate drops monotonically and vanishes at a given temperature, for physical quark masses. There is  a substantial reduction with respect to ChPT in the transition temperature, estimated here as  the vanishing condensate point, when all the hadron degrees of freedom are included. The value obtained from the plot in Fig.\ref{fig:condsushrg}  is $T_c\simeq $ 178.5 MeV, while the value obtained for instance in ChPT with three-loop  pion interactions \cite{Gerber:1988tt}, or in the virial approach \cite{GomezNicola:2012uc}, is around $T_c\simeq 250$ MeV. We will actually see in the next section that allowing some uncertainty in the normalization of the HRG expressions, to account in a simple way for the different uncertainties involved, allows for a fairly good description of lattice data.

As for the scalar susceptibility, the HRG approach showed here (and not calculated before to the best of our knowledge) gives rise to a monotonically increased function, just as  ChPT or virial approaches \cite{Gerber:1988tt,GomezNicola:2012uc}, i.e, not reproducing the transition crossover peak. Also in the next section, we will carefully explore to what extent the HRG approach can describe simultaneously the quark condensate and susceptibility lattice results.

 \section{Fits to lattice data}
\label{sec:fits}

In this section we will perform a more detailed analysis of the description of lattice data within the theoretical framework developed in this work. We will concentrate mostly in the scalar susceptibility, since, as explained before, this is the thermodynamic observable for which the role of the thermal $f_0(500)$ is expected to be more important. In particular, we will compare the description provided by thermal $f_0(500)$ saturation approach with that of the HRG, in a more quantitative way. 

As an effective way to parametrize the uncertainties in both approaches, we will allow for a normalization constant which we will consider as our fit parameter. Thus, in the thermal $f_0(500)$ saturation definition \eqref{susunit}, we fit the $A$ parameter instead of fixing it to its ChPT value, which accounts at least partially for the uncertainties inherent of this method and discussed in section \ref{sec:uchpt}.  The results we have obtained in section \ref{sec:uchpt} show that we could alternatively fit the LEC within their $T=0$ uncertainties to get a good description of lattice points, especially around the transition peak.  As for  the HRG approach, we normalize $z\rightarrow B z$ in \eqref{zHRG} as a simple way to parametrize the uncertainties in this approach such as  the quark mass dependence of hadron masses, the upper limit of the resonances included or the absence of interactions and decay channels. 

\begin{figure}
\centerline{\includegraphics[width=9cm]{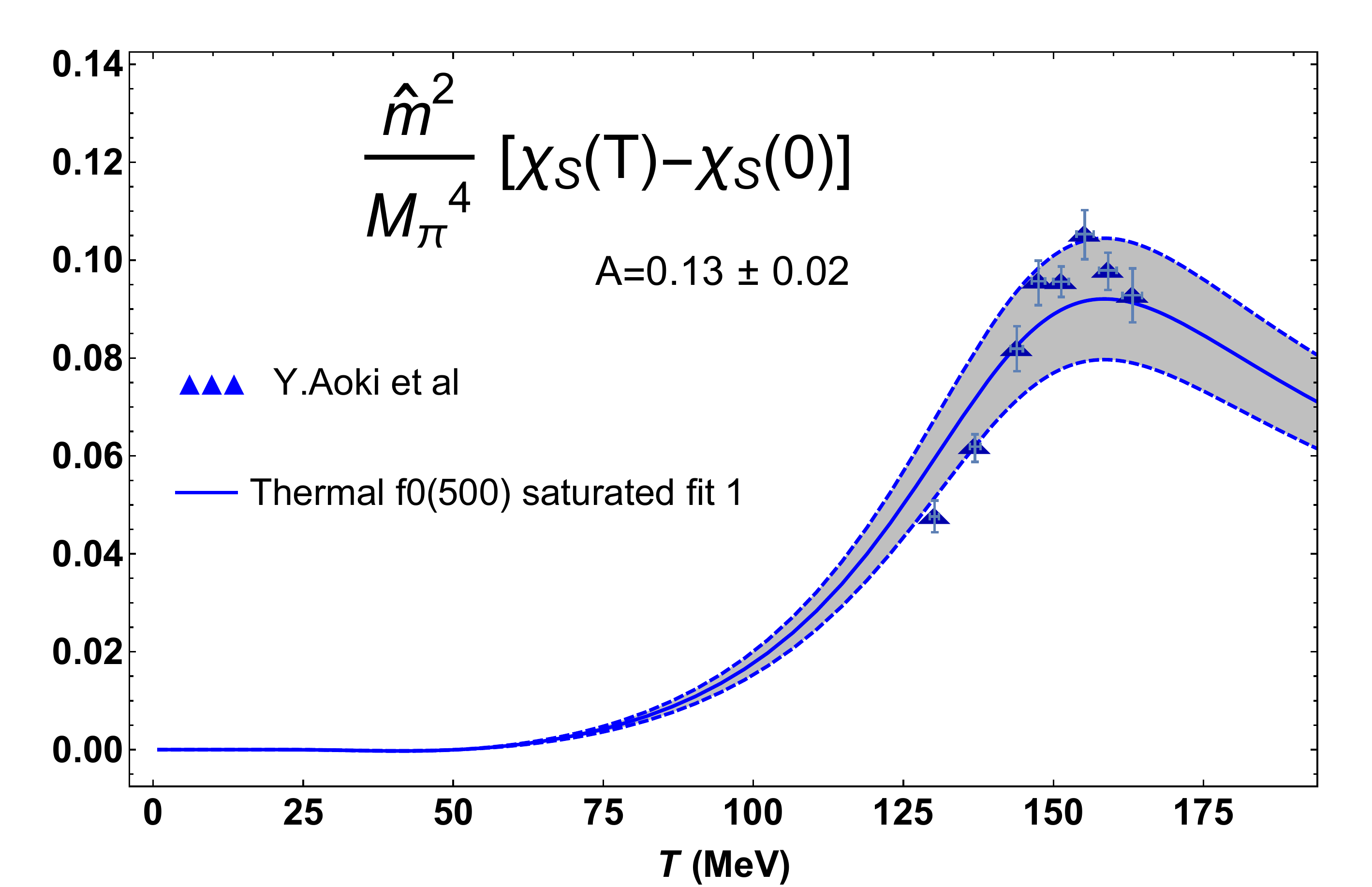}
\includegraphics[width=9cm]{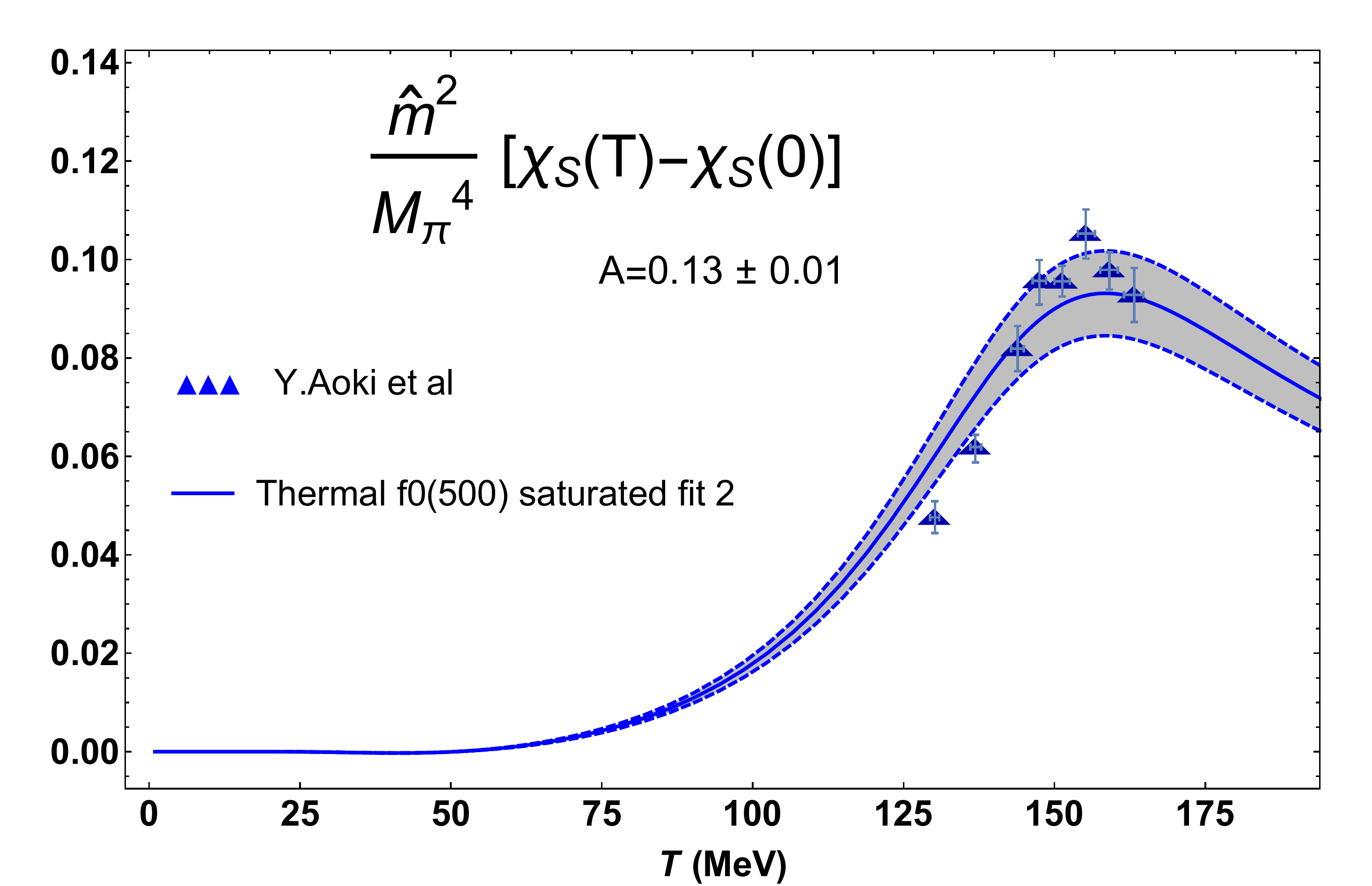}}
\caption{Fits of the thermal $f_0(500)$ saturated scalar susceptibility with the normalization constant as fit parameter and with the central values of the LEC given in \eqref{lecs}. Fit 1 corresponds to fitting data up to $T\leq T_c=155$ MeV while in fit 2 we include two more lattice points, up to $T=163$ MeV. The quoted uncertainties in the $A$ parameter and the bands correspond to the 95\% confidence level of the fit. The lattice data and errors are from \cite{Aoki:2009sc}.}
\label{fig:fitsus}
\end{figure}

We show in Fig.\ref{fig:fitsus} the results of two different fits of the thermal $f_0(500)$ saturated approach. The difference between those two fits is just the number of points included. Thus, in  fit 2 we include two more points around the transition point. The result for the $A$ parameter is shown in the figure, together with the uncertainty band corresponding to the 95\% confidence level of the fit. The different fit parameters, as well as the goodness of fit indicators are collected in Table \ref{table:fits} for all the fits performed in this section.  Note that the values of $A$ quoted in the table are compatible with the ChPT value in \eqref{Achpt}, and therefore the predictions of the fitted curve for lattice data do not spoil the expected $T=0$ value for the scalar susceptibility,  as given by  the ChPT result.

\begin{table}
\begin{tabular}{||l|c|c|c|c|c||}  \hhline{|=|=|=|=|=|=|}
	Fit & A&B&$\chi^2/$dof & $R^2$ &$T_{max}$ (MeV) \\ \hline
	Thermal $f_0(500)$  fit 1 & 0.13$\pm$ 0.02&  &6.25 & 0.986&155 \\ 
	Thermal $f_0(500)$  fit 2& 0.13$\pm$ 0.01&  & 4.93 & 0.989&165 \\
	HRG fit 1&   & 1.90$\pm$ 0.02   &1.33 &0.997&155\\
	HRG fit 2&    &  1.71$\pm$ 0.23  &10.30 &0.978&165\\
	HRG fit 3 &   & 1.06$\pm$ 0.12 & 3.77&0.998&155
	\\ \hhline{|=|=|=|=|=|=|}	\end{tabular}
\caption{Parameters for the different fits as explained in the main text.}
\label{table:fits}
\end{table}

\begin{figure}
\centerline{\includegraphics[width=9cm]{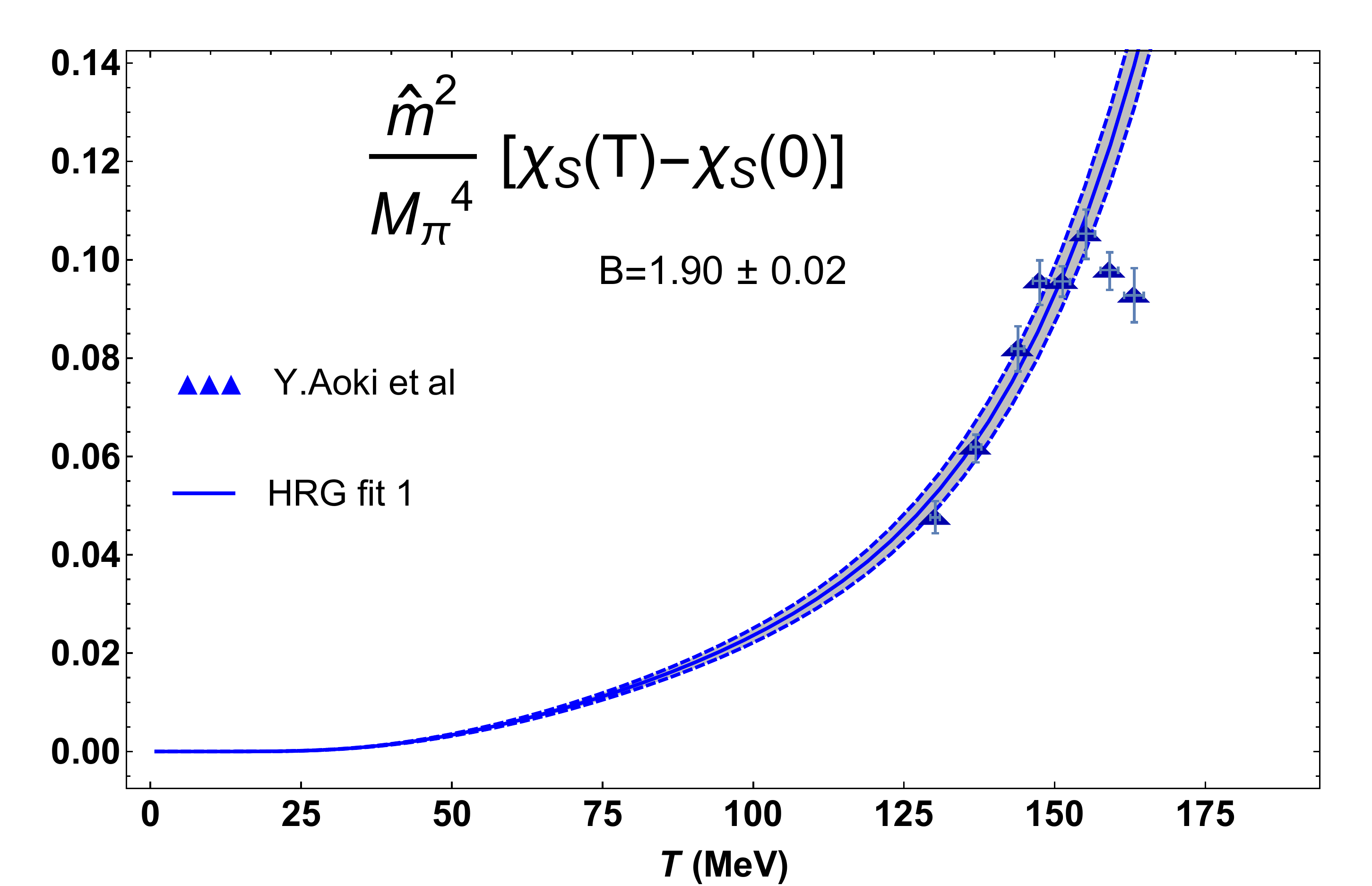}
\includegraphics[width=9cm]{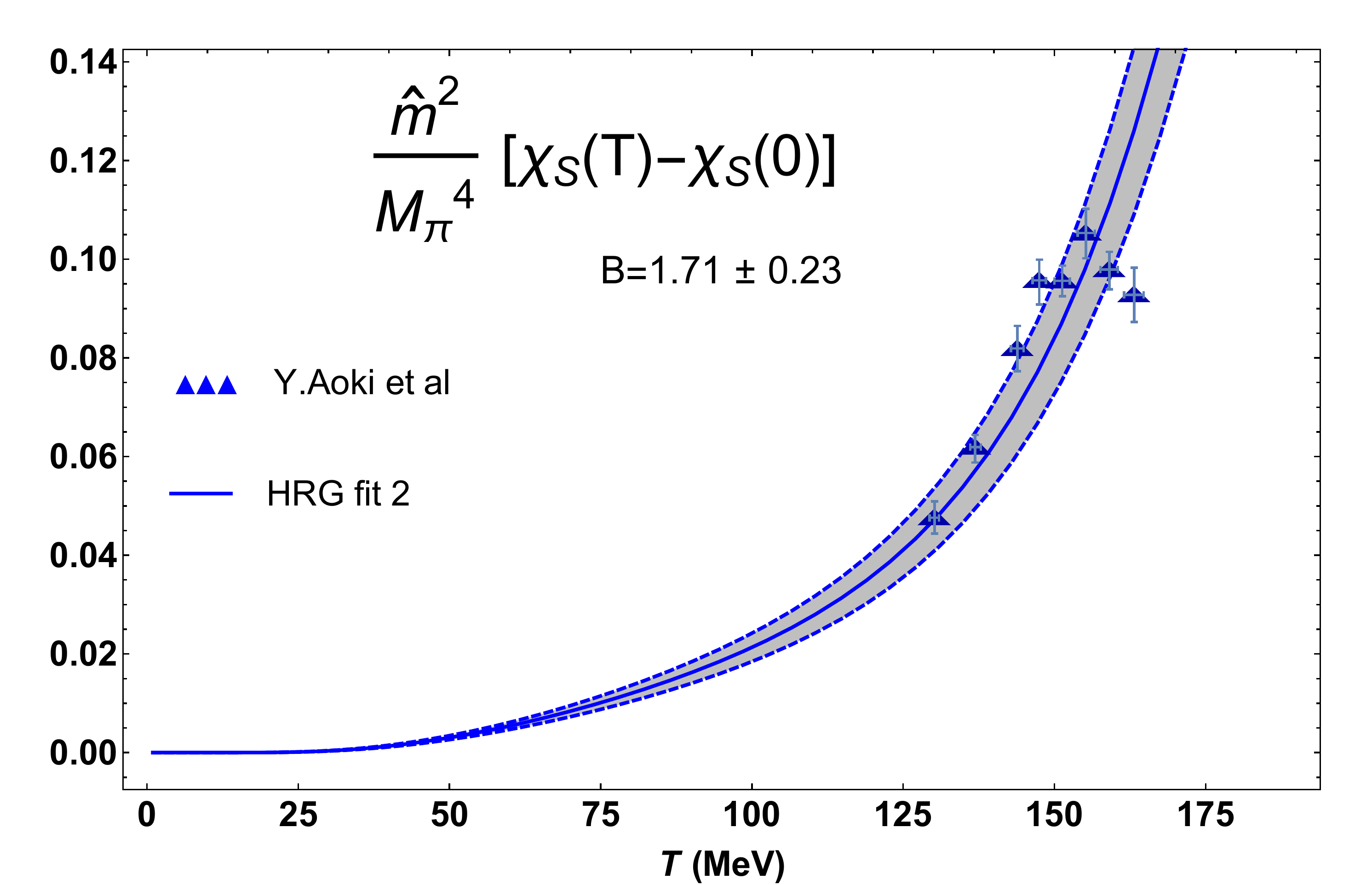}}
\caption{Fits of the HRG scalar susceptibility with the normalization constant as fit parameter. Fit 1 corresponds to fitting data up to $T\leq T_c=155$ MeV while in fit 2 we include two more lattice points, up to $T=163$ MeV. The quoted uncertainties in the $B$ parameter and the bands correspond to the 95\% confidence level of the fit. The lattice data and errors are from \cite{Aoki:2009sc}.}
\label{fig:fithrg12}
\end{figure}

On the other hand, in Fig.\ref{fig:fithrg12} we show the results of two fits with the HRG approach (HRG fits 1 and 2),  corresponding to fit only the susceptibility lattice points, with the same sets of data used for fits 1 and 2 with the thermal $f_0(500)$ approach. We see that, as long as we keep the data points below $T_c$, the HRG gives a slightly better fit than the thermal $f_0(500)$ one, as it would be naturally expected from a HRG approach. However, including only two more points around $T_c$  worsens the HRG in favor of the $f_0(500)$ one, which is consistent with the different qualitative behaviour of both curves around the maximum and confirms our previous comments about the role of the thermal $f_0(500)$. 

\begin{figure}
\centerline{\includegraphics[width=9cm]{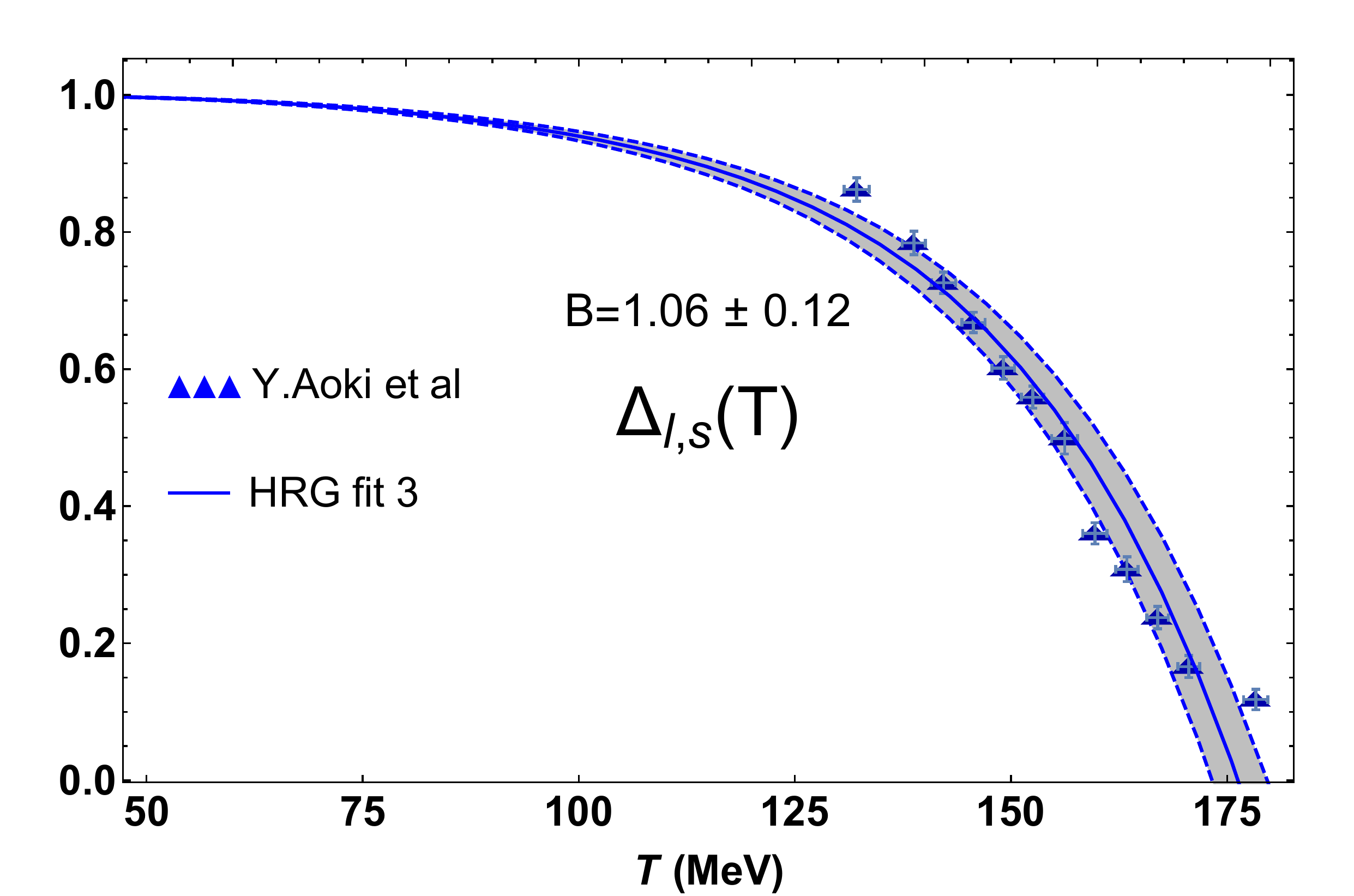}
\includegraphics[width=9cm]{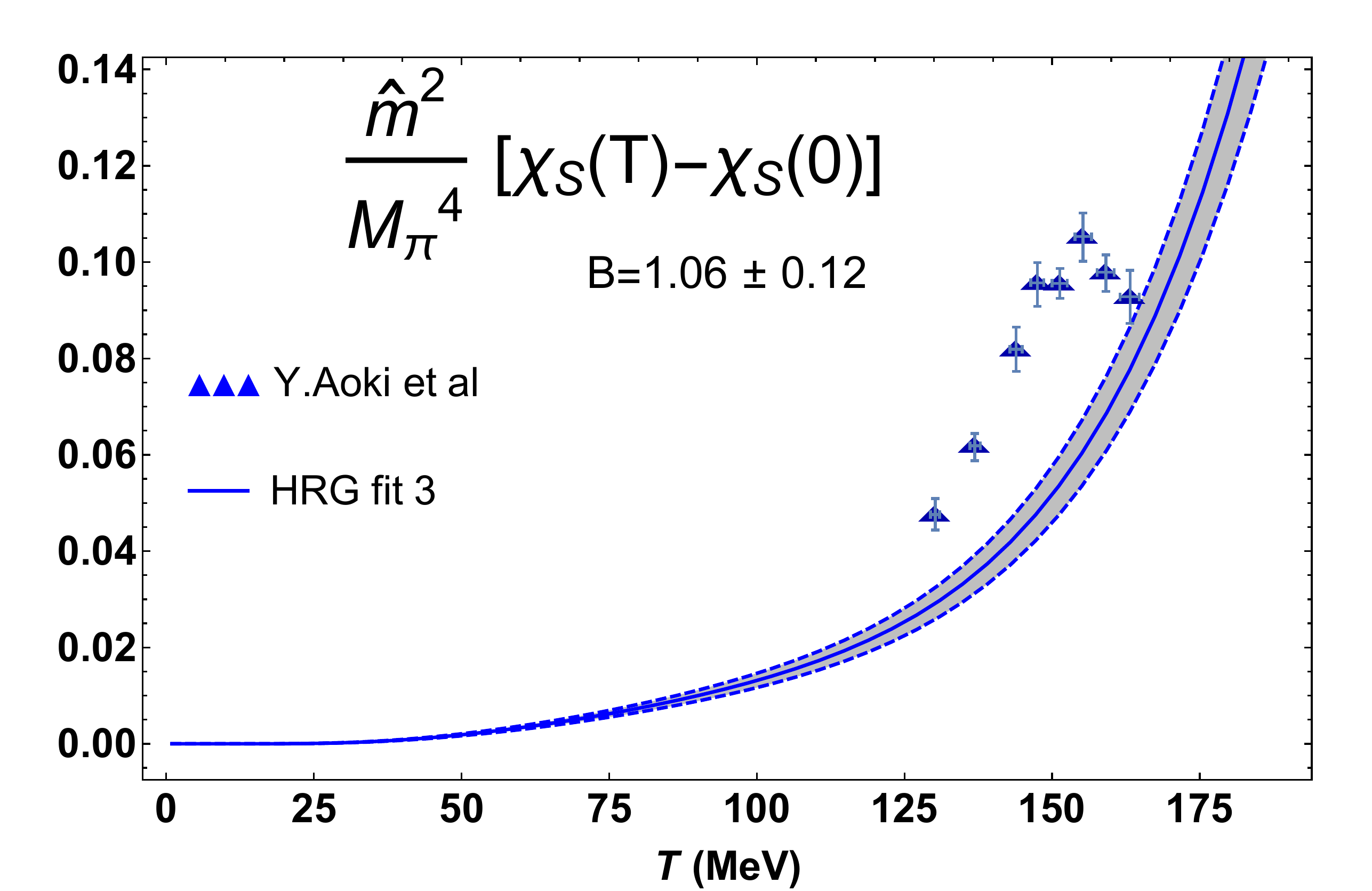}}
\caption{Left:Fit of the HRG reduced condensate with the normalization constant as fit parameter, fitting data up to $T\leq T_c=155$ MeV. The quoted uncertainties in the $B$ parameter and the bands correspond to the 95\% confidence level of the fit. The lattice data and errors are from \cite{Aoki:2009sc}. Right: prediction for the scalar susceptibility with the same $B$ central value and uncertainty.}
\label{fig:fithrg3}
\end{figure}

Regarding the HRG description, an important observation must be taken into account: the values of $B$ needed to fit the susceptibility are in conflict with those needed to fit the quark condensate. Let us justify  this conclusion in detail. For that purpose, we consider  the HRG result for the  reduced quark condensate

\begin{equation}
\Delta_{l,s}=\frac{\condl (T)-\frac{m_l}{m_s}\conds(T)}{\condl (0)-\frac{m_l}{m_s}\conds(0)},
\end{equation}
which is one of the condensate combinations for which lattice analysis yield definite predictions, being free of finite-size divergences~\cite{Aoki:2009sc,Borsanyi:2010bp,Bazavov:2011nk,Buchoff:2013nra}. We include the fitting $B$ parameter as explained before 
 (recall that  $B$  multiplies only the finite temperature correction of quark condensates, not their $T=0$ part). The result of such fit (HRG fit 3) is provided in Fig.\ref{fig:fithrg3}, the fit parameters being  given in Table \ref{table:fits}, and shows a very good description of the reduced condensate, with a value of $B$ compatible with unity and therefore in agreement with the analysis in \cite{Jankowski:2012ms}. However, that value is incompatible with that in fit 2, i.e,  the HRG scalar susceptibility fit in the same temperature range. Such incompatibility is clearly seen in  the prediction for $\chi_S$ showed in the right panel of that figure. Recall that those lattice data for both quantities come exactly from the same collaboration and lattice setup. The previous claim is confirmed if we try to fit jointly the reduced condensate and scalar susceptibility lattice points. In that case, we obtain a $\chi^2/\mbox{dof} \simeq 71$ indicating clearly that such a joint description of both quantities within the HRG approach is not feasible.

We could of course perform more elaborated fits, such as considering the LEC in the thermal $f_0(500)$ or the hadron masses and their quark mass dependence in the HRG as additional fit parameters. However, the main objective of our present analysis is to compare both approaches and show that actually the thermal $f_0(500)$ one is competitive with respect to the HRG around the transition, and a simple one-parameter fit is enough for such purposes.

 \section{Conclusions}
 \label{sec:conc}

In this work we have performed a detailed  analysis of the importance of the thermal corrections to the $f_0(500)$ resonance spectral parameters, regarding the description of the scalar susceptibility $\chi_S$ around the region of chiral symmetry restoration. Such analysis has been carried out for different realizations of the thermal $f_0(500)$ state within effective theories.  First, using the LSM as a testbed, we have showed that a direct relation can be established between the scalar susceptibility and the propagator of the lightest scalar state at zero momentum.  Through the analysis of the LSM one-loop $\sigma$ self-energy at finite temperature, we have shown that the susceptibility saturated by the $\sigma$ propagator has a much larger growth than the purely perturbative one, approaching better lattice data, although with a divergent behaviour in the massive case.  The LSM analysis provides additional support for the formulation of $\chi_S$ through the UChPT saturated approach, where the $f_0(500)$ arises as a resonance in $\pi\pi$ scattering, including thermal corrections. The UChPT approach provides a much more reliable description of the $T=0$ $f_0(500)$ pole and of $\chi_S (T)$ as long as the basic requirements of unitarity and analiticity are maintained.  Within the IAM formulation, such approach actually reproduces correctly the crossover peak and lattice data within the sensitivity of the ChPT low-energy constants. The requirements of unitarity, analiticity and a good determination of the $T=0$ pole are crucial  to achieve the expected qualitative behaviour for the thermal scalar mass, although a correct description of the saturated susceptibility is achieved when the full $\Od(p^4)$  corrections to the thermal amplitude are taken into account.  

A  conclusion shared by the LSM and UChPT  approaches  is that a saturated approach for $\chi_S(T)$ where only this thermal state is included,  can account for most of lattice data below and even around the transition. For that reason, we have performed several fits of the UChPT saturated approach, with a single parameter fit (normalization factor) comparing it with a description based on the Hadron Resonance Gas where all hadron states below 2 GeV have been included. The HRG result for $\chi_S(T)$, which had not been analyzed before, provides a better fit  than UChPT for temperatures below the transition. However, as  values closer to $T_c$ are included, the UChPT improves over the HRG, since it can describe the susceptibility peak. In addition, the HRG fits for the scalar susceptibility are in conflict with those of the quark condensate, using a single-parameter fit. 

Through the various approaches analyzed in this work, we conclude that the  thermal $f_0(500)$ state is crucial to describe correctly the scalar susceptibility and hence to understand correctly the chiral restoration transition.  We believe that our results can be useful in that sense and we leave for future work related problems such as the possibility to include thermal interactions for the scalar channel within the HRG which could help to understand previous studies regarding the role of the $f_0(500)$ in that approach.

 \section*{Acknowledgments}
 We are very grateful to J. Ruiz de Elvira and J. Sanz-Cillero  for useful comments and discussions.  
Work partially supported by  research contract FPA2016-75654-C2-2-P  (spanish ``Ministerio de Econom\'{\i}a y Competitividad").

\end{document}